\documentclass[%
 reprint,
superscriptaddress,
 amsmath,amssymb,
 aps,
]{revtex4-2}

\usepackage{graphicx}% Include figure files
\usepackage{dcolumn}% Align table columns on decimal point
\usepackage{bm}% bold math

\usepackage{url}
\usepackage[colorlinks=true, linkcolor=blue, citecolor=blue, urlcolor=blue, breaklinks=true]{hyperref}

\begin{document}

\preprint{APS/123-QED}

\title{Emergence, Evolution and Manipulation of Swing Voters in Presidential Election}% Force line breaks with \\

\author{Ziqian Liu}
\affiliation{School of Artificial Intelligence, Beihang University, Beijing 100191, China}
\affiliation{Beijing Advanced Innovation Center for Future Blockchain and Privacy Computing, Beihang University, Beijing 100191, China}
\affiliation{Key laboratory of Mathematics, Informatics and Behavioral Semantics, Beihang University, Beijing 100191, China}

\author{Xin Wang}
\email{wangxin\_1993@buaa.edu.cn}
\affiliation{School of Artificial Intelligence, Beihang University, Beijing 100191, China}
\affiliation{Beijing Advanced Innovation Center for Future Blockchain and Privacy Computing, Beihang University, Beijing 100191, China}
\affiliation{Key laboratory of Mathematics, Informatics and Behavioral Semantics, Beihang University, Beijing 100191, China}
\affiliation{Zhongguancun Laboratory, Beijing 100094, China}
\affiliation{State Key Laboratory of Complex \& Critical Software Environment, Beihang University, Beijing 100191, China}

\author{Junyu Lu}
\affiliation{School of Artificial Intelligence, Beihang University, Beijing 100191, China}
\affiliation{Key laboratory of Mathematics, Informatics and Behavioral Semantics, Beihang University, Beijing 100191, China}

\author{Longzhao Liu}
\affiliation{School of Artificial Intelligence, Beihang University, Beijing 100191, China}
\affiliation{Beijing Advanced Innovation Center for Future Blockchain and Privacy Computing, Beihang University, Beijing 100191, China}
\affiliation{Key laboratory of Mathematics, Informatics and Behavioral Semantics, Beihang University, Beijing 100191, China}
\affiliation{Zhongguancun Laboratory, Beijing 100094, China}
\affiliation{State Key Laboratory of Complex \& Critical Software Environment, Beihang University, Beijing 100191, China}

\author{Hongwei Zheng}
% \email{hwzheng@pku.edu.cn}
\affiliation{Beijing Academy of Blockchain and Edge Computing, Beijing 100085, China}

\author{Shaoting Tang}
\email{tangshaoting@buaa.edu.cn}
\affiliation{School of Artificial Intelligence, Beihang University, Beijing 100191, China}
\affiliation{Beijing Advanced Innovation Center for Future Blockchain and Privacy Computing, Beihang University, Beijing 100191, China}
\affiliation{Key laboratory of Mathematics, Informatics and Behavioral Semantics, Beihang University, Beijing 100191, China}
\affiliation{Zhongguancun Laboratory, Beijing 100094, China}
\affiliation{State Key Laboratory of Complex \& Critical Software Environment, Beihang University, Beijing 100191, China}
\affiliation{Hangzhou International Innovation Institute, Beihang University, Hangzhou 311115, China}
\affiliation{Institute of Trustworthy Artificial Intelligence, Zhejiang Normal University, Hangzhou 310012, China}
\affiliation{Institute of Medical Artificial Intelligence, Binzhou Medical University, Yantai 264003, China}

% \date{\today}% It is always \today, today,
             %  but any date may be explicitly specified

\begin{abstract}
Political polarization, fueled by public discourse and echo chambers, threatens the foundation of democratic elections. However, traditional one-dimensional opinion models---assuming ``support for one party equals opposition to another''---fail to capture the nuanced dynamics of swing voters (including neutrals, left leaners and right leaners), who are critical for the final election outcomes. This study introduces a two-dimensional opinion model that classifies voters into five groups, enabling precise characterization of the swing group's interactive behaviors. Importantly, we introduce antagonism effect to describe the intensities with which the two camps incite opposition and exert voting pressure in the run-up to the election, typically via Us-versus-Them framing. By integrating the open-mindedness of voters, the stubbornness of opinion interactions, and the antagonism effect manipulated by the two parties, we systematically explore the intricate interplay between top-down political campaigns and bottom-up interpersonal opinion dynamics, unveiling their nonlinear coupling impacts on the emergence, and evolution of swing voters.
Counterintuitively, we find that extreme antagonism effects might backfire in presidential election: when both parties adopt intense antagonistic strategies, the party that polarizes more strongly risks alienating swing voters, thereby enabling its ostensibly weaker opponent to prevail. These insights are also validated on the core retweet networks during 2020 U.S. presidential election. Building upon multidimensional opinion model, our results highlight the possibility of manipulating swing voters and shaping electoral outcomes through antagonistic strategies of political parties. Our work also provides a nuanced and generalizable framework for analyzing opinion dynamics in other polarized public discourse.
\end{abstract}

%\keywords{Suggested keywords}%Use showkeys class option if keyword
                              %display desired
\maketitle

%\tableofcontents

\section{INTRODUCTION}

Political polarization has emerged as a widespread feature of contemporary democratic politics, with empirical research consistently documenting its deepening roots in online information ecosystems \cite{kubin2021role}. 
Substantial volumes of online network data afford scholars invaluable insights into the investigation of political polarization across diverse sociopolitical contexts, such as the US presidential elections \cite{flamino2023political,gonzalez2023asymmetric}, the impeachment of the former Brazilian President Dilma Roussef \cite{cota2019quantifying}, COVID-19 pandemic \cite{jungkunz2021political,kerr2021political}, vaccination \cite{johnson2020online} and environmental protection \cite{falkenberg2022growing}.
Homophilous interactions force individuals to increasingly engage with like-minded peers, leading to the formation of echo chambers and information cocoons \cite{cinelli2021echo,interian2023network}.
Conversely, polarized information ecosystems restructure social networks via information cascades, causing users to lose cross-ideological ties at rates exceeding random chance \cite{tokita2021polarized}.
Empirical analyses further highlight how algorithmic recommendation systems amplify ideological segregation \cite{santos2021link}, while the spread of false news reveals that misinformation disproportionately reaches and mobilizes partisan groups, marginalizing the middle ground \cite{vosoughi2018spread,grinberg2019fake}.

Amidst this backdrop of escalating political polarization, the swing groups---comprising neutrals and leaners---emerge as pivotal yet paradoxical actors in democratic discourse \cite{mayer2008swing,kosmidis2010undecided,mahieux2024investigation,mathur2022information,schill2017angry}. 
The importance of swing groups resides in their dual role as bridges for cross-ideological dialogue and critical decision-maker of political mobilization. 
Empirical analyses reveal that these groups serve as ``bridge builders'' in networked environments, maintaining connectivity between polarized camps and mediating cross-ideological information flow \cite{cox200913,throsby2018deciders}.
Moreover, their positional centrality in networks like Twitter during election debates highlights their capacity to shift political trends \cite{siegenfeld2020negative,pratelli2024online}.
Yet this importance is matched by profound complexity which stems from their multidimensional attitudinal configurations and context-dependent behaviors \cite{falkenberg2022growing,baumann2021emergence,kwon2024configurations,andrade2021multidimensional}.
Especially in the context of presidential elections, the diversity of political issues (e.g., immigration, tax reform and education) and the complexity of partisan ideologies create conditions for ideological uncertainty among swing voters \cite{feldman2014understanding,walgrave2013ideology,hamill1985breadth}.

Despite the empirical evidence highlighting the significant role of swing groups, modeling frameworks historically oversimplify their complexity, often treating them as homogeneous ``undecided'' blocs rather than dynamic, multidimensional actors \cite{wang2020public}. As an effective modeling approach, opinion dynamics plays an important role in theoretically understanding how microscopic interaction rules among agents give rise to macroscopic patterns of group phenomena, such as consensus, polarization or fragmentation \cite{galesic2019statistical,nowak2019nonlinear}. 
Based on the classic DeGroot model of opinion dynamics which assumes that individuals update their opinions by taking weighted averaging opinions of their networked neighbors \cite{baronchelli2018emergence,degroot1974reaching,french1956formal}, a large class of extensions of this mechanism have been proposed by introducing additional assumptions, such as Friedkin-Johnse (FJ) model with initial prejudice \cite{friedkin1990social} and Hegselmann–Krause (HK) model with bounded-confidence \cite{deffuant2000mixing}.
In social networks, homophilic interactions and preferential engagement with like-minded peers have given rise to another novel class of models designed to capture the escalation of opinion polarization and the emergence of echo chambers \cite{baumann2020modeling,baumann2021emergence}.

Existing research on opinion dynamics has predominantly relied on one-dimensional continuous models, which represent attitudes as points on a linear spectrum. While these frameworks effectively capture simple polarization and fragmentation, they constrained by the implicit binary-opposition hypothesis that support for one party is assumed to inherent opposition to the other \cite{rahmawati2009dynamics,herdin2012deconstructing,martinek2007right}, exhibiting fundamental limitations in capturing the complexity of swing group evolution. 
Furthermore, the top-down dynamics---such as elite-driven strategies \cite{rothschild2020elites,li2023collective}, algorithm curation \cite{liu2021interaction,cho2020search,ibrahim2023youtube}, and media framing \cite{entman2010media}---remain underexplored in modeling swing group evolution and manipulation, despite their pervasive impact.

Therefore, this paper focuses on more precisely modeling the swing group and studying how its evolution and the manipulation strategies of political parties affect the final election outcomes.
We develop a mesoscale multi-cognitive model based on two-dimensional opinions, which explicitly incorporates open-mindedness of voters \cite{pansanella2022change,schawe2020open,dolbier2024open}, the stubbornness of opinion interactions \cite{ghaderi2014opinion}, and the antagonism effect from external political campaigns to capture the emergence, evolution and manipulation of swing voters.
First, by separately considering the different views of voters towards the two parties, we assign each voter a two-dimensional opinion vector and categorize voters into five groups---Party A supporters,  Party A leaners,  neutrals, Party B leaners, and Party B supporters---providing a more nuanced representation of the voter population \cite{mayer2007swing,hill2017changing,chang2023active}. We then integrate this multi-cognitive framework with traditional models to describe a cross-scale group opinion dynamics. 
Our results show the dual role for antagonism in political system: while it enhances voter mobilization, it simultaneously erodes the diversity of ideological interaction and fosters two-dimensional echo chambers, trapping voters' opinion updating within two homogeneous networks.
Our model also reveals the complex interaction between the top-down antagonistic political campaign and bottom-up individual stubbornness, triggering the emergence and evolution of neutral groups. Counterintuitively, competitive antagonism between parties has a complex impact on electoral outcomes: under weak to moderate antagonism, the party with more intense antagonistic strategies secures a voting advantage by mobilizing more swing voters; while with both parties deploying extreme antagonism, the party with weaker antagonism gains more support from neutrals, leading to a reversal of voting advantage. Finally, we validate our proposed model on 2020 U.S. election retweet networks and demonstrate its robustness in real world.

\section{MODEL}
We now first introduce this multi-cognitive framework based on the two-dimensional coupling opinions to achieve the measurement and characterization of multiple cognitive groups including party supporters, party leaners and the neutrals (see schematic in Figure ~\ref{fig:1}(a)).

\subsection{Agent Properties}
Each voter is characterized by a two-dimensional opinion vector $o_i(t)=(x_i(t),y_i(t))\in[0,1]^2$, where $x_i(t)$ and $y_i(t)$ represent the cognitions towards Party A and Party B respectively at time $t$. A clear preference for Party A corresponds to $x_i\approx1$ and $y_i\approx0$, while strong support for Party B results in $y_i\approx1$ and $x_i\approx0$.
The opinion difference $z_i=x_i-y_i$ serves as the attitudinal indicator, whose symbolic value can measure the attitude and voting tendency of the voter $i$ towards the two parties. A positive $z_i(t)$ ($sign(z_i)=1$) indicates Party A preferences, whereas a negative value ($sign(z_i)=-1$) reflects Party B inclination.

To further characterize the complexity of political spectrum, especially the differences within swing voters, we divide the population into five types based on $z_i$ and two thresholds $z_o$ (outer boundary for strong partisans) and $z_v$ (inner boundary for neutrals): Party A supporters ($z_i(t){>}z_o$), Party A leaners ($z_v{<}z_i(t){\leq} z_o$), neutrals ($|z_i(t)|{\leq}z_v$), Party B leaners ($-z_o{\leq} z_i(t){<}{-}z_v$), and Party B supporters ($z_i(t){<}{-}z_o$). Among them, leaners and neutrals compose the swing voter population. 
The interactive classification parameter $\sigma_o(i,t)$ for opinion dynamics is defined as:
\begin{equation}
    \sigma_o(i,t)=\begin{cases}
    A, & z_{i}(t){>}z_{o} \\
C, & z_{o}{\geq} z_{i}(t){\geq}{-}z_{o} \\
B, & z_{i}(t){<}{-}z_{o}
    \end{cases}
\end{equation}
Here, A,B,C respectively designate Party A supporters, Party B supporters and swing voters.

The evolution of the opinions of the voters ends with the vote in the election, and the total duration of the interaction and the update of the opinions is recorded as $T$.
It is noteworthy that at election time $T$, only neutrals ($|z_i(T)|{\leq} z_v$) abstain from voting, while leaners and supporters cast votes according to their final $z_i(T)$.

\subsection{Network Structure}
The model employs a graph $G=(V,E)$ with $N$ nodes representing voters. Each node $v_i$ features  a self-loop $(v_i,v_i)\in E$, enabling self-opinion updating independent of external interactions. The edge $(v_i, v_j)\in E$ represents the path of interaction between the voter $i$ and the neighboring voter $j$.

Based on their current state $\sigma_o(j,t)$, neighbor nodes $v_j$ of given $v_i$ are dynamically categorized into three mesoscale interaction groups:

(1) $v_j$ is the Party A supporter. The sets of $x$-opinion and $y$-opinion neighbors for $v_i$ at time $t$ are as follows:
\begin{equation}
    \left\{
\begin{aligned}
\mathcal{N}_{i}^{AX}(t)=&\{j|(v_j, v_i)\in E, \sigma_{o}(j, t)=A, \\ &|x_i(t)-x_j(t)|<\epsilon\}  \\
\mathcal{N}_{i}^{AY}(t)=&\{j|(v_j, v_i)\in E, \sigma_{o}(j, t)=A, \\&|y_i(t)-y_j(t)|<\epsilon\}
\end{aligned}
\right.
\end{equation}

(2) $v_j$ is the Party B supporter. The sets of $x$-opinion and $y$-opinion neighbors for $v_i$ at time $t$ are as follows:
\begin{equation}
\left\{
\begin{aligned}
\mathcal{N}_{i}^{BX}(t)=&\{j|(v_j, v_i)\in E, \sigma_{o}(j, t)=B, \\ & |x_i(t)-x_j(t)|<\epsilon\} \\
\mathcal{N}_{i}^{BY}(t)=&\{j|(v_j, v_i)\in E, \sigma_{o}(j, t)=B,  \\& |y_i(t)-y_j(t)|<\epsilon\}
\end{aligned}
\right.
\end{equation}

(3) $v_j$ is the swing voter. The sets of $x$-opinion and $y$-opinion neighbors for $v_i$ at time $t$ are as follows:
\begin{equation}
    \left\{
\begin{aligned}
\mathcal{N}_{i}^{CX}(t)=&\{j|(v_j, v_i)\in E, \sigma_{o}(j, t)=C,  \\& |x_i(t)-x_j(t)|<\epsilon\}  \\
\mathcal{N}_{i}^{CY}(t)=&\{j|(v_j, v_i)\in E, \sigma_{o}(j, t)=C,  \\&  |y_i(t)-y_j(t)|<\epsilon\}
\end{aligned}
\right.
\end{equation}
Here, the open-mindedness $\epsilon$ represents the interaction threshold between nodes and their neighbors. Only when their opinions are close enough will they interact and influence each other.

\subsection{Updating Rules}
\begin{figure*}[htbp]
\centering
\includegraphics[width=\textwidth]{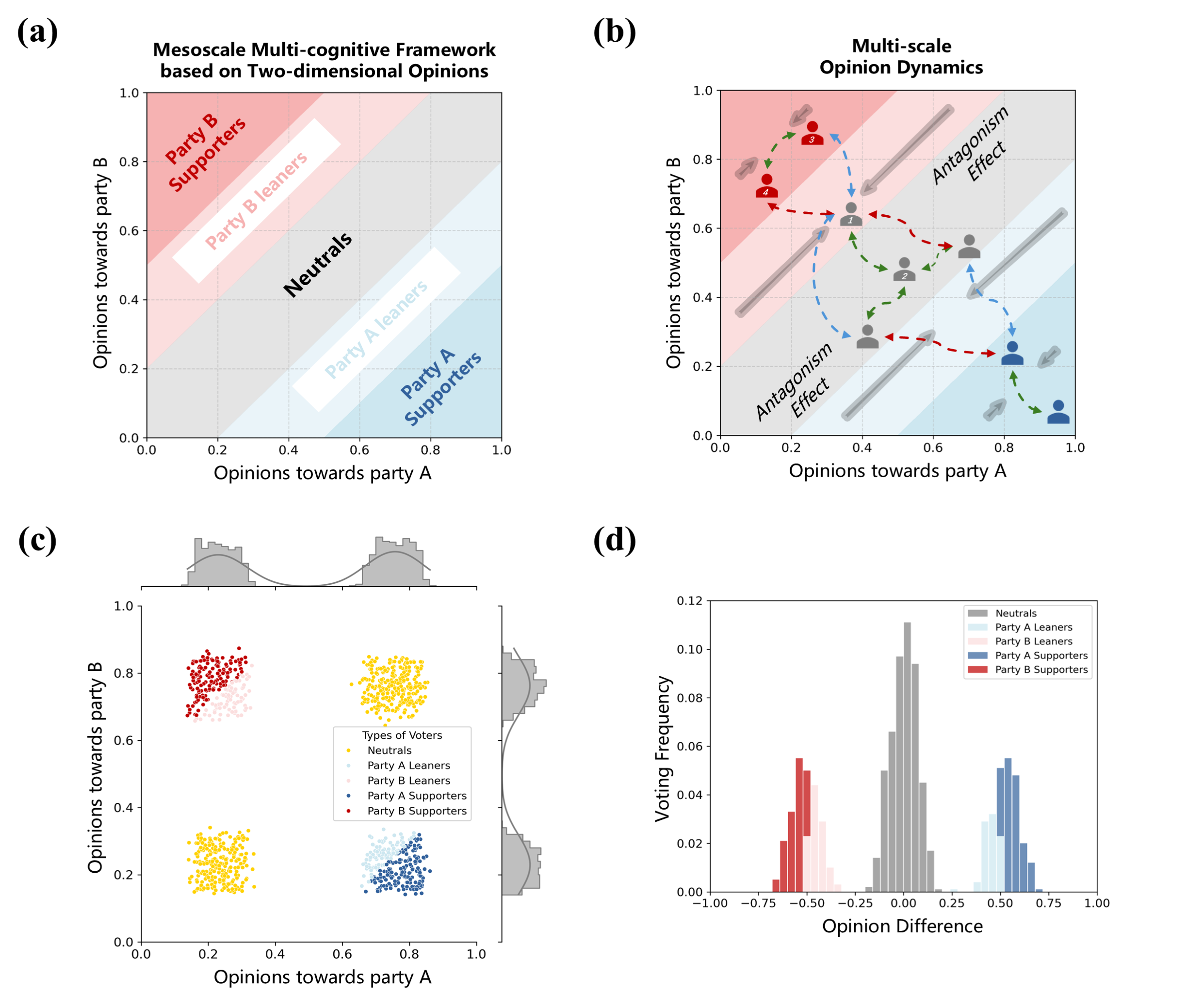}% Here is how to import EPS art
\caption{\label{fig:1} (a) Schematic of the mesoscale multi-cognitive framework based on two-dimensional coupling opinions. Voters are divided into five categories:  Party A supporters, Party A leaners, neutrals, Party B leaners, and Party B supporters according to the disparities of their opinions towards Party A and B. (b) Schematic of the multi-scale opinion dynamics which incorporates voters' initial preferences, network interactions and antagonism effect. The blue, red, and green arrows represent the interactions between opinions towards Party A, Party B, and both parties, respectively. Among them, voter 1 interacts with voter 3 and voter 4 respectively on opinions towards Party A and Party B, while voter 1 and voter 2 interact with each other on opinions towards both parties. Furthermore, the antagonism effect forces the opinions of all voters to converge in the direction of one-dimensional opposition. (c-d) An example of distributions of two-dimensional coupling opinions and voting outcomes. Polarization emerges in the group opinions of the two parties, and the voting outcomes show a trimodal distribution, with the neutral non-voting group appearing. Parameters: the example begins from an ER graph with $N=10^3$ and $\langle k \rangle=40$, $z_o=0.5$, and $z_v=0.2$. The fixed parameters are: $\epsilon=0.2$, $\lambda=0.3$, $\rho=0$.}
\end{figure*}

Building upon the mesoscale multi-cognitive framework, we now establish a novel multi-scale opinion dynamics mechanism (see schematic in Figure ~\ref{fig:1}(b)).
This model explicitly incorporates the open-mindedness of voters, the stubbornness of opinion interactions, and importantly, the antagonism effect manipulated by external political campaigns to capture the realistic dynamics of partisan competition and swing voter mobilization in presidential elections.

The evolution of opinions balances the initial preferences and neighbor interactions. In the absence of antagonism, the heterogeneous updating rules for node $v_i$ are defined as:

(1) Party A supporters

Voters identified as Party A supporters update their $x$-opinions by interacting with Party A supporters and swing voters whose $x$-opinions lie within the interaction threshold, and update their $y$-opinions by interacting with Party A supporters and swing voters whose $y$-opinions lie within the interaction threshold.
\begin{equation}
    \left\{
\begin{aligned}
\hat{x}_i(t+1) &= \lambda_{i} x_i(0) \\
&+ (1 - \lambda_{i}) \bigg( w_{AX} \frac{1}{|\mathcal{N}_i^{AX}(t)|} \sum_{j \in \mathcal{N}_i^{AX}(t)} x_j(t) \\
&+ w_{CX} \frac{1}{|\mathcal{N}_i^{CX}(t)|} \sum_{k \in \mathcal{N}_i^{CX}(t)} x_k(t) \bigg) \\
\hat{y}_i(t+1) &= \lambda_{i} y_i(0)\\
&+ (1 - \lambda_{i}) \bigg( w_{AY} \frac{1}{|\mathcal{N}_i^{AY}(t)|} \sum_{j \in \mathcal{N}_i^{AY}(t)} y_j(t) \\
&+ w_{CY} \frac{1}{|\mathcal{N}_i^{CY}(t)|} \sum_{k \in \mathcal{N}_i^{CY}(t)} y_k(t) \bigg)
\end{aligned}
\right.
\end{equation}

(2) Party B supporters

Voters identified as Party B supporters update their $x$-opinions by interacting with Party B supporters and swing voters whose $x$-opinions lie within the interaction threshold, and update their $y$-opinions by interacting with Party B supporters and swing voters whose $y$-opinions lie within the interaction threshold.
\begin{equation}
    \left\{
\begin{aligned}
\hat{x}_i(t+1) &= \lambda_{i} x_i(0)\\
&+ (1 - \lambda_{i}) \bigg( w_{BX} \frac{1}{|N_i^{BX}(t)|} \sum_{j \in N_i^{BX}(t)} x_j(t) \\
&+ w_{CX} \frac{1}{|N_i^{CX}(t)|} \sum_{k \in N_i^{CX}(t)} x_k(t) \bigg) \\
\hat{y}_i(t+1) &= \lambda_{i} y_i(0)\\
&+ (1 - \lambda_{i}) \bigg( w_{BY} \frac{1}{|N_i^{BY}(t)|} \sum_{j \in N_i^{BY}(t)} y_j(t) \\
&+ w_{CY} \frac{1}{|N_i^{CY}(t)|} \sum_{k \in N_i^{CY}(t)} y_k(t) \bigg)
\end{aligned}
\right.
\end{equation}

(3) Swing voters

Voters identified as swing voters update their $x$-opinions by interacting with all neighbors whose $x$-opinions lie within the interaction threshold and update their $y$-opinions by interacting with all neighbors whose $y$-opinions lie within the interaction threshold.
\begin{equation}
    \left\{
\begin{aligned}
\hat{x}_i(t+1) &= \lambda_{i} x_i(0) \\
&+ (1 - \lambda_{i}) \bigg( w_{AX} \frac{1}{|\mathcal{N}_i^{AX}(t)|} \sum_{j \in \mathcal{N}_i^{AX}(t)} x_j(t) \\
&+ w_{BX} \frac{1}{|\mathcal{N}_i^{BX}(t)|} \sum_{k \in \mathcal{N}_i^{BX}(t)} x_k(t) \\
&+ w_{CX} \frac{1}{|\mathcal{N}_i^{CX}(t)|} \sum_{l \in \mathcal{N}_i^{CX}(t)} x_l(t) \bigg) \\
\hat{y}_i(t+1) &= \lambda_{i} y_i(0)\\
&+ (1 - \lambda_{i}) \bigg( w_{AY} \frac{1}{|\mathcal{N}_i^{AY}(t)|} \sum_{j \in \mathcal{N}_i^{AY}(t)} y_j(t) \\
&+ w_{BY} \frac{1}{|\mathcal{N}_i^{BY}(t)|} \sum_{k \in \mathcal{N}_i^{BY}(t)} y_k(t) \\
&+ w_{CY} \frac{1}{|\mathcal{N}_i^{CY}(t)|} \sum_{l \in \mathcal{N}_i^{CY}(t)} y_l(t) \bigg)
\end{aligned}
\right.
\end{equation}

Opinions of voters evolve through independent updates for each party dimension, incorporating contributions from initial preferences, where voters retain inherent stubbornness quantified by parameter $\lambda\in[0,1]$ that controls the weight of initial opinions $x_i(0),y_i(0)$, and network interactions, which involve weighted averages of mesoscale group opinions with the strength of intergroup influence being quantified by the weight $w$.

%The self-loop ensures non-empty neighbor sets for all nodes. For node $i: \sigma_o(i,t)=A$, $\mathcal{N}_i^{AA}(t)\ne\varnothing$ and $\mathcal{N}_i^{AB}(t)\ne\varnothing$. For node $i: \sigma_o(i,t)=B$, $\mathcal{N}_i^{BA}(t)\ne\varnothing$ and $\mathcal{N}_i^{BB}(t)\ne\varnothing$. For node $i: \sigma_o(i,t)=C$, $\mathcal{N}_i^{CA}(t)\ne\varnothing$ and $\mathcal{N}_i^{CB}(t)\ne\varnothing$. These configurations ensure well-defined cognitive dynamics, with valid weight assignments detailed in Table~\ref{tab:table1},~\ref{tab:table2},~\ref{tab:table3}. For additional cases of assignments, refer to Appendix~\ref{app: 3}.
In the main text, we fix all the influence weight for simplicity (see Table~\ref{tab:table1},~\ref{tab:table2},~\ref{tab:table3}) and further verify their robustness in Appendix~\ref{app: 3}.

\renewcommand{\arraystretch}{1.3}
\begin{table}[htbp]
\centering
\caption{\label{tab:table1} possible influence weight for $i$: $\sigma_{o}=A$}
\begin{tabular}{p{0.3\columnwidth} p{0.15\columnwidth} p{0.15\columnwidth} p{0.15\columnwidth} p{0.15\columnwidth}}
\hline
Influence Weight & $w_{AX}$ & $w_{CX}$ & $w_{AY}$ & $w_{CY}$\\
\hline
If $\mathcal{N}_{i}^{CX}=\varnothing$ & 1 & 0 & & \\

If $\mathcal{N}_{i}^{CX}\neq\varnothing$ & 0.6 & 0.4 & &\\

If $\mathcal{N}_{i}^{CY}=\varnothing$ & & & 1 & 0 \\

If $\mathcal{N}_{i}^{CY}\neq\varnothing$& & & 0.6 & 0.4 \\
\hline
\end{tabular}
\end{table}

\begin{table}[htbp]
\centering
\caption{\label{tab:table2}possible influence weights for $i$: $\sigma_{o}=B$}
\begin{tabular}{p{0.3\columnwidth} p{0.15\columnwidth} p{0.15\columnwidth} p{0.15\columnwidth} p{0.15\columnwidth}}
\hline
Influence Weight & $w_{BX}$ & $w_{CX}$ & $w_{BY}$ & $w_{CY}$ \\
\hline
If $\mathcal{N}_{i}^{CX}=\varnothing$ & 1 & 0 & & \\

If $\mathcal{N}_{i}^{CX}\neq\varnothing$ & 0.6 & 0.4 & & \\

If $\mathcal{N}_{i}^{CY}=\varnothing$ & & & 1 & 0 \\

If $\mathcal{N}_{i}^{CY}\neq\varnothing$ & & & 0.6 & 0.4 \\
\hline
\end{tabular}
\end{table}

\begin{table}[htbp]
\centering
\caption{\label{tab:table3}possible influence weights for $i$: $\sigma_{o}=C$}
\begin{tabular}{l l l l l l l}
\hline
Influence Weight & \(w_{AX}\) & \(w_{BX}\) & \(w_{CX}\) & \(w_{AY}\) & \(w_{BY}\) & \(w_{CY}\) \\
\hline
If \(\mathcal{N}_{i}^{AX}=\varnothing\) and \(\mathcal{N}_{i}^{BX}=\varnothing\) & 0 & 0 & 1 & & & \\

If \(\mathcal{N}_{i}^{AX}=\varnothing\) and \(\mathcal{N}_{i}^{BX}\neq\varnothing\) & 0 & 0.5 & 0.5 & & & \\

If \(\mathcal{N}_{i}^{AX}\neq\varnothing\) and \(\mathcal{N}_{i}^{BX}=\varnothing\) & 0.5 & 0 & 0.5 & & & \\

If \(\mathcal{N}_{i}^{AX}\neq\varnothing\) and \(\mathcal{N}_{i}^{BX}\neq\varnothing\) & 0.3 & 0.3 & 0.4 & & & \\

If \(\mathcal{N}_{i}^{AY}=\varnothing\) and \(\mathcal{N}_{i}^{BY}=\varnothing\) & & & & 0 & 0 & 1 \\

If \(\mathcal{N}_{i}^{AY}=\varnothing\) and \(\mathcal{N}_{i}^{BY}\neq\varnothing\) & & & & 0 & 0.5 & 0.5 \\

If \(\mathcal{N}_{i}^{AY}\neq\varnothing\) and \(\mathcal{N}_{i}^{BY}=\varnothing\) & & & & 0.5 & 0 & 0.5 \\

If \(\mathcal{N}_{i}^{AY}\neq\varnothing\) and \(\mathcal{N}_{i}^{BY}\neq\varnothing\) & & & & 0.3 & 0.3 & 0.4 \\
\hline
\end{tabular}
\end{table}

In addition, the stubbornness of all voters are equally quantified by $\lambda_i=\lambda\in[0,1]$, thus the opinion dynamics can be expressed as:

\begin{equation}
\label{eq:000}
    \left\{
\begin{aligned}
\hat{x}_i(t+1) &= \lambda x_i(0) + (1 - \lambda) \bar{x}_i(t) \\
\hat{y}_i(t+1) &= \lambda y_i(0) + (1 - \lambda) \bar{y}_i(t)
\end{aligned}
\right.
\end{equation}
Here $\bar{x}_i(t),\bar{y}_i(t)$ denote weighted means of $x-$opinions and $y-$opinions from neighboring voters, respectively. 

The opinion updating rule, mathematically equivalent to a convex combination of initial and neighbor-averaged opinions, can be reformulated as a convex optimization problem seeking the global minimum of the following objective function:
%\begin{equation}
%    \left\{
%\begin{aligned}
%x_i(t+1) &= \mathop{\arg \min}_{x_i} \lambda \cdot (x_i - x_i(0))^2 + (1 - \lambda)(x_i - %\bar{x}_i(t))^2 \\
%y_i(t+1) &= \mathop{\arg\min}_{y_i} \lambda \cdot (y_i - y_i(0))^2 + (1 - \lambda)(y_i - %\bar{y}_i(t))^2
%\end{aligned}
%\right.
%\end{equation}
%and merge into:

%x_i(t+1), y_i(t+1) &= \mathop{\arg \min}_{x_i,y_i}

\begin{equation}
\begin{aligned}
  \hat{\mathcal{L}}(x_i,y_i) &= \lambda (x_i - x_i(0))^2 + (1 - \lambda)(x_i - \bar{x}_i(t))^2 \\
  &+\lambda (y_i - y_i(0))^2 + (1 - \lambda)(y_i - \bar{y}_i(t))^2
\end{aligned}
\end{equation}

Further, to describe the intensities with which the two camps incite opposition and exert voting pressure in the run-up to the election, we introduce the antagonism effect that couples the two-dimensional opinion dynamics, via zero-sum framing. The resulting function is as follows:
\begin{equation}
\begin{aligned}
\mathcal{L}(x_i,y_i) &=  (1 - \rho_i) \big[ \lambda (x_i - x_i(0))^2 + (1 - \lambda)(x_i - \bar{x}_i(t))^2 \\
&+ \lambda (y_i - y_i(0))^2 + (1 - \lambda)(y_i - \bar{y}_i(t))^2 \big] \\ &+\rho_i (x_i + y_i - 1)^2
\end{aligned}
\label{eq:111}
\end{equation}
Eq.\eqref{eq:111} displays the complex interplay of three key factors: initial partisan preferences, mesoscale-mediated interactions and antagonism coupling between party opinions where $\rho_i\in[0,1]$ tunes the strengths of antagonism effect.

Intuitively, when $\rho=1$, the antagonism reaches its maximum, which means the binary hypothesis holds strictly ($x_i+y_i=1$), and the formula degenerates into a one-dimensional case; while when $\rho=0$, the formula decomposes into two independent updating rules of classical opinion dynamics in Eq.\eqref{eq:000}.

Solving the optimization problem yields the following explicit expression for opinion dynamics:
\begin{equation}
    \left\{
\begin{aligned}
x_i(t+1) &= \frac{\hat{x}_i(t+1) - \rho_i \hat{y}_i(t+1) + \rho_i}{1 + \rho_i} \\
y_i(t+1) &= \frac{\hat{y}_i(t+1) - \rho_i \hat{x}_i(t+1) + \rho_i}{1 + \rho_i}
\end{aligned}
\right.
\end{equation}
The detailed convexity proof and the solution process for the objective function are provided in the Appendix~\ref{app: 1}. The PYTHON implementation of the model simulations is available \cite{liu2025github}.

Without considering the antagonism effect ($\rho=0$), Figures.~\ref{fig:1}(c) and (d) respectively illustrate examples of the two-dimensional opinion distribution and the voting outcomes of the voter groups emerging at the macro level. Under moderate open-mindedness, the group opinions towards both parties exhibit polarization, which is consistent with classic opinion dynamics; but the final voting behavior presents a three-peak distribution. Besides the voting groups of the two parties, there are a large number of neutrals that do not participate in the voting. The opinions of this non-voting community towards two parties are quite similar. They may hold both supportive or both opposing opinions simultaneously, reproducing the complexity of swing groups that cannot be described by traditional cognitive models based on one-dimensional opinions and binary hypothesis.

\section{RESULTS}

\subsection{Opinion dynamics under varying open-mindedness}

In the scenario where no antagonism effect is operative, the opinion-updating processes of the two parties unfold independently. Herein, the degree of open-mindedness plays a crucial role in shaping the evolution of opinions.

\begin{figure}[htbp]
\centering
\includegraphics[width=\linewidth]{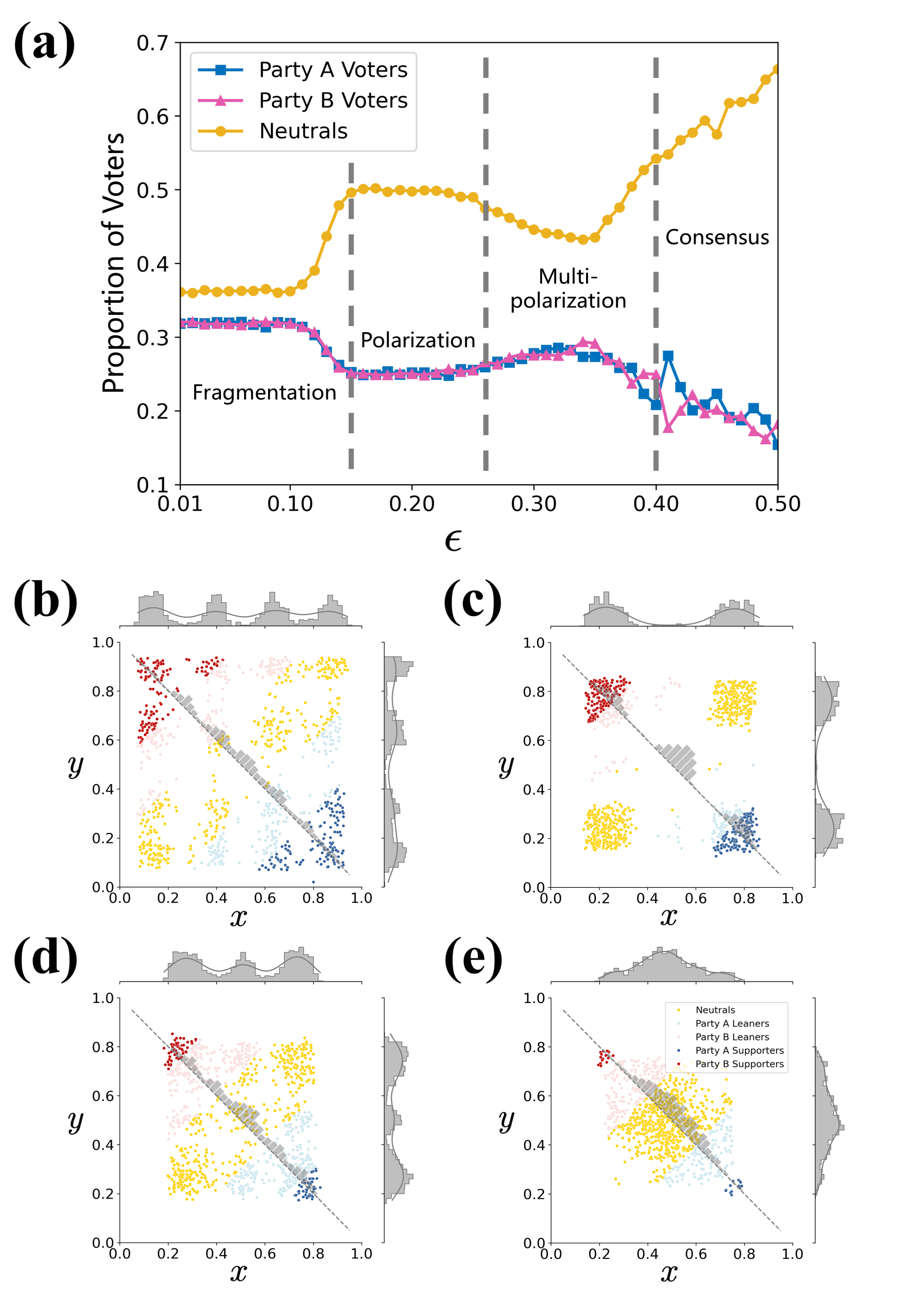}% Here is how to import EPS art
\caption{\label{fig:2}
Open-mindedness and opinion evolution. (a) The open-mindedness (characterized by the interaction threshold) of voters significantly influences the distribution of two-dimensional opinions. (b) Small open-mindedness lead to opinion segmentation. (c) Under a moderate level of open-mindedness, the ideological stances of both parties emerge polarization. (d,e) The continuously enhanced open-mindedness goes through multipolarization and gradually fosters a broad consensus around neutrality.
Simulation results are averaged over 100 independent runs. Parameters: simulations in (a) begin from an ER graph with $N=10^4$ and $\langle k \rangle=40$, $z_o=0.5$, $z_v=0.2$, $\rho=0$, and $\lambda=0.3$. We change (b) $\epsilon=0.1$. (c) $\epsilon=0.25$. (d) $\epsilon=0.35$. (e) $\epsilon=0.45$.
}
\end{figure}

We begin from the exploration of the distribution of group opinions and voting outcomes across a spectrum of open-mindedness degrees. For a given open-mindedness, we not only characterize the two-dimensional distribution of group opinions but also analyze the marginal distributions of partisan attitudes. 
We explore the number of opinion clusters by employing the kernel density estimation (KDE) curve of the $x$-opinion and $y$-opinion distributions and computing the number of its local maxima \cite{davis2011remarks,silverman2018density,scott2015multivariate}.
We utilize the average number of $x$-opinion and $y$-opinion peaks to represent the number of opinion clusters.
Taking polarization (with a cluster number of 2) as the boundary, we classify the opinion clusters under different open-mindedness into four distinct regimes: fragmentation, polarization, multipolarization, and consensus, and present the voting outcomes in Figure~\ref{fig:2}(a).

At the extreme end of the spectrum, when $\epsilon$ is exceedingly low, the opinions of both parties exhibit a multi-clustered ideological partitions, while the ideological stances of both parties gradually converge to a neutral position with a exceedingly high $\epsilon$. 
Under a low open-mindedness, voters are highly resistant to different ideas, leading to the formation of variegated opinion distribution and balanced proportion of neutrals and actual voters (Fig.~\ref{fig:2}(b)). Conversely, under a particularly high open-mindedness, the free flow of ideas and willingness of voters to consider alternative perspectives facilitate the formation of broad consensus and the emergence of a dominating neutral group (Fig.~\ref{fig:2}(e)).

When the degree of open-mindedness is at a moderate level, the well-known phenomenon of ideological polarization emerges distinctly within each party (Fig.~\ref{fig:2}(c)). Consequently, the voting outcomes exhibit distinct tripolar characteristics with a significant proportion of the neutral group.
However, the transition from polarized group opinion to neutral consensus involves an intermediate state of multipolarization (Fig.~\ref{fig:2}(d)). Under a wider open-mindedness, some swing voters absorb political discourses from both parties, leading to the emergence of multiple scattered swing blocs. Notably, the multipolarization of partisan opinions also results in diversified voting outcomes, as the spread of ideological subgroups encourages broader electoral participation.

During competitive presidential elections, the opinion distributions of fragmentation and neutral consensus are unlikely to occur. Therefore we focus on the dynamics of group polarization, with particular emphasis on the evolution and strategic manipulation of swing groups. In the subsequent investigation, we constrain the parameter of open-mindedness within a moderate regime.

\subsection{Voting polarization and two-dimensional echo chambers under antagonism}

\begin{figure*}[htbp]
\centering
\includegraphics[width=\textwidth]{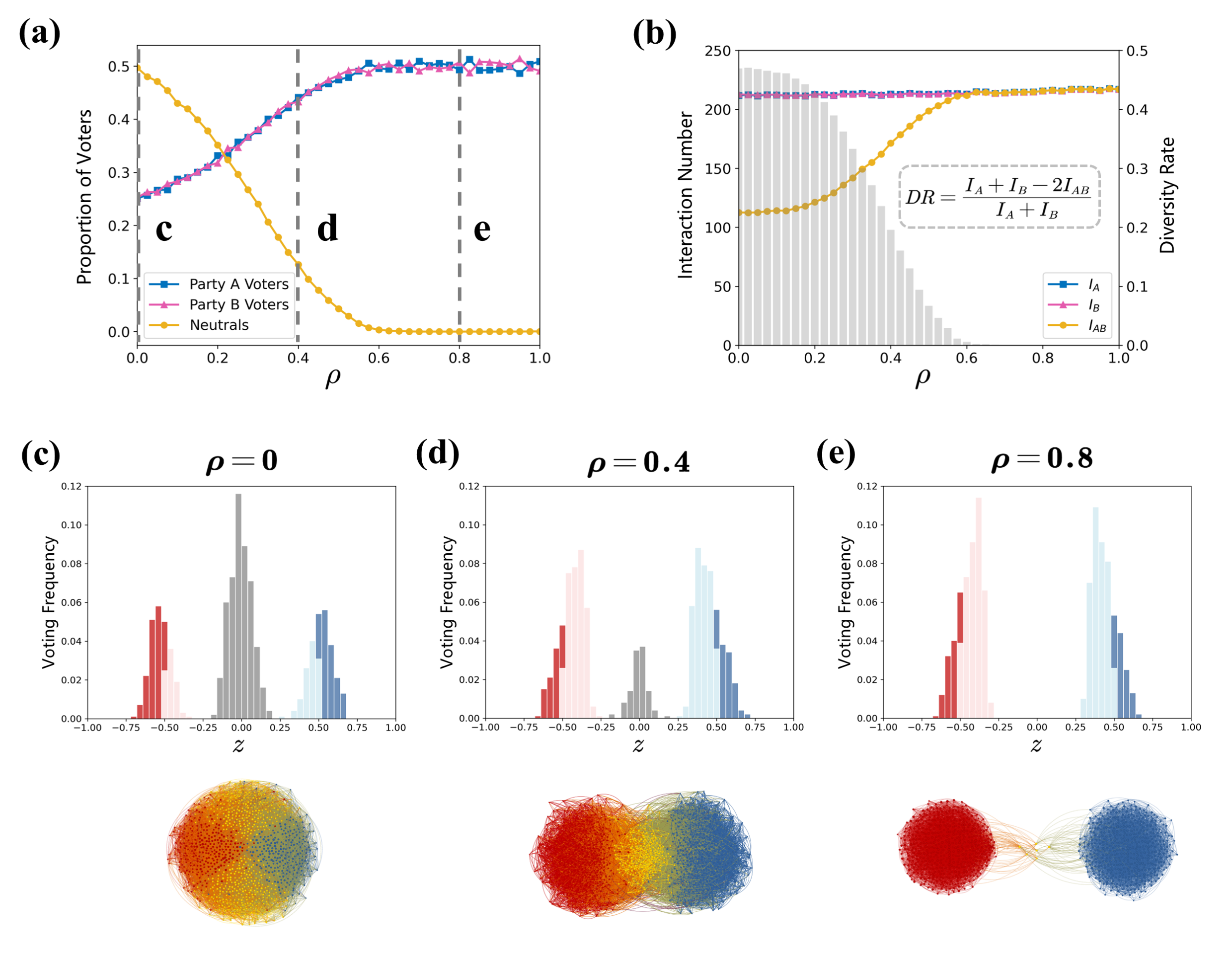}% Here is how to import EPS art
\caption{\label{fig:3}
The emergence of two-dimensional echo chambers under antagonism effect. Increasing the level of antagonism are sufficient to (a) promote the voting participation of swing voters and (b) reduce the diversity of opinion interactions.
In (b), $I_A$, $I_B$ and $I_{AB}$ respectively denote the average number of interacting neighbors regarding $x-$opinion, $y-$opinion, and both opinions. 
(c)-(e) show stable voting distributions and network structures corresponding to different strength of antagonism effects, illustrating the elimination of the neutrals and the emergence of two-dimensional echo chambers. Parameters: simulations begin from an ER graph with $N=10^4$ and $\langle k \rangle=40$, $ z_o=0.5$, $z_v=0.2$, $\epsilon=0.2$, and $\lambda=0.3$. We change (c) $\rho=0$. (d) $\rho=0.4$. (e) $\rho=0.8$.
}
\end{figure*}

At the moderate level of open-mindedness in the absence of antagonism effects, the ideological stances of both parties exhibit pronounced polarization, which turn into tripolarization in voting behavior. This state arises from the independent evolution of each party's opinion-updating system, unmoderated by cross-ideological interaction or external pressure. The neutral non-voting group reflects the ideological ambivalence or strategic abstention of voters between the two polarized extremes.

In this section, we incorporate the antagonism effect that couples the two-dimensional opinions and explore how such external partisan rivalry reshapes opinion dynamics and voting behavior through mobilizing previously inactive voters. Here, all voters are subject to a homogeneous level of antagonism, allowing us to evaluate how the competition between two parties influences swing voter behaviors.

As the antagonism effect becomes stronger, Figure~\ref{fig:3}(a) presents a clear trend in voting outcomes: the proportion of neutrals decreases monotonically, while the voting rates of both parties increase. This reflects the mobilizing effect of heightened partisan rivalry, which incentivizes swing voters---particularly neutrals---to commit to a certain party rather than remain inactive. Under extremely strong antagonism, the electorate reaches full participation, with voting outcomes exhibiting complete polarization where no neutral bloc remains. This transition underscores the role of antagonism as a catalyst for electoral engagement.

To quantify the degree of social polarization of the coupled two-dimensional opinions, we introduce the diversity rate ($DR$) defined as $DR=(I_A+I_B-2I_{AB})/(I_A+I_B)$, where $I_A$, $I_B$ and $I_{AB}$ denote the average number of interacting neighbors regarding $x$-opinion, $y$-opinion and both opinions, respectively. Intuitively, $DR=0$ means voters' perception of both parties are almost entirely shaped by the same interaction networks, indicating the emergence of two-dimensional echo chambers where the diversity of opinion interaction reaches its lowest point. When $DR=1$, in the contrary, the interaction networks of both parties are totally independent.% which is inconsistent with real political situations.

In Figure~\ref{fig:3}(b) we show that $DR$ also decreases monotonically. At low antagonism, over 40\% of information transmission remains party-specific, allowing independent ideological exposure. As antagonism levels exceeding 0.5, $DR$ approaches zero and this homogenization of opinion interaction suppresses cross-ideological exchange, driving the formation of isolated echo chambers where voters are increasingly insulated from opposing views. 
These insights can also be observed from the stable voting distributions and network structures under different antagonism effects shown in Figure~\ref{fig:3}(c-e).

Our findings reveal a dual role for antagonism in political system: while it enhances voter mobilization, it simultaneously erodes the diversity of ideological interaction. The convergence of $I_A$,$I_B$, and $I_{AB}$ under high antagonism demonstrates how partisan competition can create self-reinforcing feedback loops, where voters' opinions become trapped within homogeneous networks---a hallmark of the two-dimensional echo chamber.

\begin{figure*}[htbp]
\centering
\includegraphics[width=\textwidth]{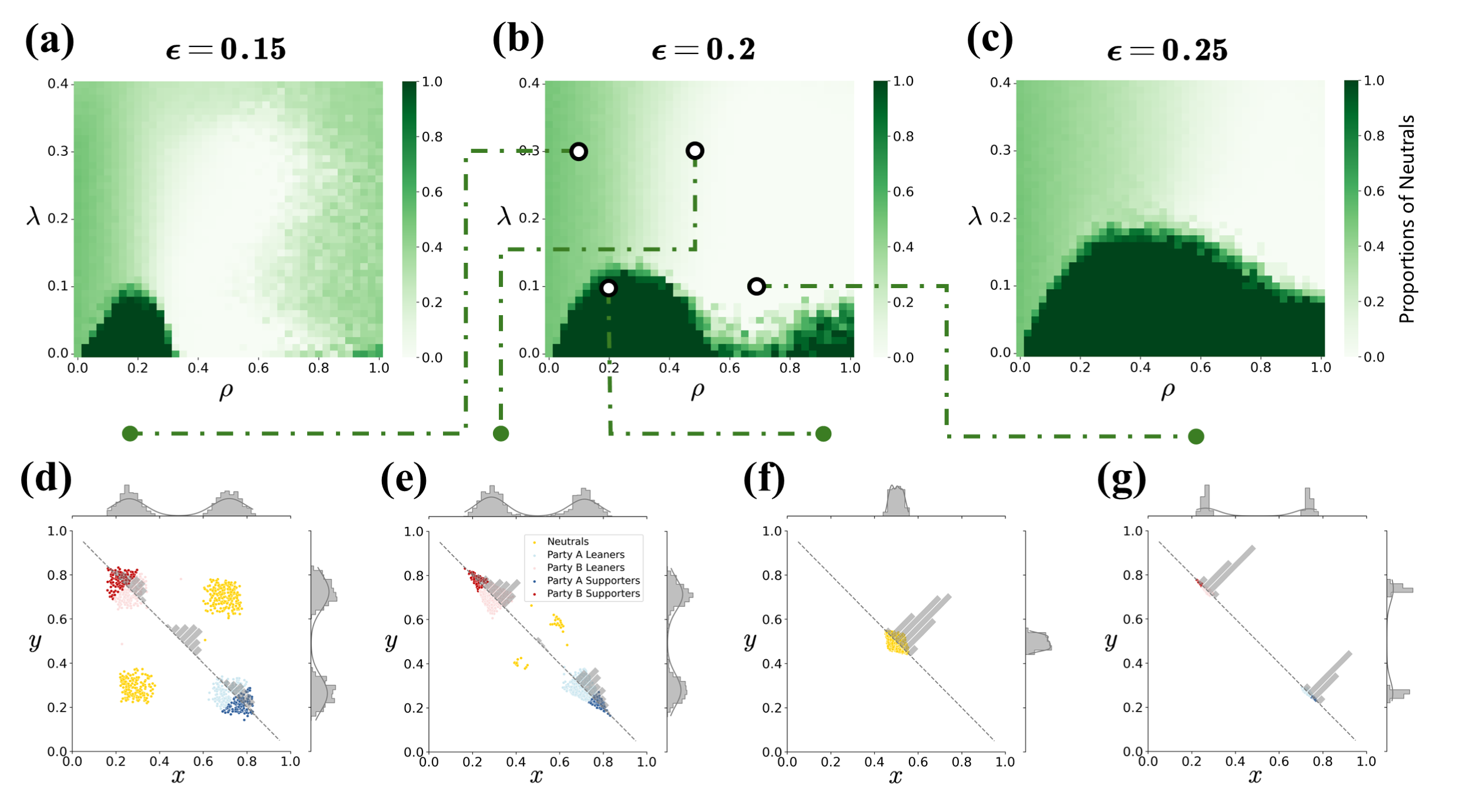}% Here is how to import EPS art
\caption{\label{fig:4}
Nonlinear coupling between stubbornness and antagonism effect: evolution of the swing group. We show the number of neutrals at low (a, $\epsilon$=0.15), moderate (b, $\epsilon$=0.2), and high (c, $\epsilon$=0.25) open-mindedness. A relatively high value of stubbornness maintains the polarization of partisan opinions, while low stubbornness can cause complex phase transitions between polarization and consensus. (d-g)We present the distribution of two-dimensional opinions and voting outcomes under four typical cases. Simulation results are averaged over 50 independent runs. Parameters: simulations begin from an ER graph with $N=10^4$ and $\langle k \rangle=40$, $z_o=0.5$, and $z_v=0.2$. We change (d) $\epsilon=0.2, \lambda=0.3, \rho=0.1$. (e) $\epsilon=0.2, \lambda=0.3, \rho=0.5$. (f) $\epsilon=0.2, \lambda=0.1, \rho=0.2$. (g) $\epsilon=0.2, \lambda=0.1, \rho=0.7$.}
\end{figure*}

\subsection{Nonlinear coupling between stubbornness and antagonism}

In Figure~\ref{fig:4}, we further investigate the intricate interplay between the top-down antagonism ($\rho$) exerted by external political campaigns and the bottom-up individual stubbornness ($\lambda$) applying to opinion dynamics.

Figure~\ref{fig:4}(b) clearly illustrates the complex nonlinear influence of the coupling effects on voting behavior under moderate open-mindedness. 
For moderate and relatively high values of $\lambda$, the neutral voter bloc is small, revealing that an adequate level of individual stubbornness promotes the formation and maintenance of political polarization.
Under this circumstance, an increase in $\rho$ drives a monotonic decrease in neutral voters, which is consistent with our prior observations (Fig.~\ref{fig:4}(d),(e)). 
Nevertheless, when the group's stubbornness is extremely high, voters are more likely to adhere to their initial opinions and resistant to external antagonism. As a result, the neutral group maintains a certain size and cannot be completely eliminated.

Conversely, when $\lambda$ is relatively small, voters' opinions become highly malleable and easily swayed by external factors, making it challenging to maintain a stable polarized state. 
When $\rho$ is small, the effect of antagonism in reducing ambiguous opinions is relatively slow. Meanwhile, it provides broader groups of voters with more opportunities to communicate and interact, leading the opinion evolution to shift from polarization to consensus (Fig.~\ref{fig:4}(f)). 
As $\rho$ continues to increase, the antagonism effect redominates the coupling system, with ambiguous opinions rapidly move towards the two extremes and even stronger polarization has emerged (Fig.~\ref{fig:4}(g)).

In the extreme case, where $\lambda$ is extremely small and $\rho$ is extremely large, population consensus achieves again. This counterintuitive phenomenon arises from the interplay between the high plasticity of voters and the externally imposed pressure to take political sides. In this scenario, the two-dimensional opinions of all voters rapidly converge towards the line $x+y=1$ under strong antagonism effect. This convergence facilitates more cross-group interactions, making even extreme supporters susceptible to the opinions of leaners. Through extensive intergroup exchanges, neutral voters bridge the partisan divide, ultimately driving the system to global consensus.

For completeness, we also explore the detailed coupling effects under lower and higher open-mindedness in Figures~\ref{fig:4}(a) and (c), respectively. The overall trends in $\rho-\lambda$ phase plane largely align with the behavioral patterns in Figure~\ref{fig:4}(b).
Besides, under a larger $\epsilon$, voters are more susceptible to a wider range of interpersonal influence, driving the system to a broader consensus. In contrast, a smaller $\epsilon$ restricts large-scale communications, facilitating the emergence of stronger polarization.
% Under intense antagonism, the resulting opinion landscape retains a robust neutral group that maintains intermediate attitudes towards both parties. This persistent neutrality gives rise to tripolar voting distributions, where the neutral groups persists as a distinct behavioral cluster.

In summary, our results have underscored the complex interplay between top-down political campaign and bottom-up interpersonal opinion dynamics. The coupling of antagonism, stubbornness and open-mindedness exerts significant nonlinear effects on the emergence and evolution of swing groups, particularly the neutrals, which has a profound impact on the final election results.

\subsection{Manipulation on swing voters: heterogeneous antagonism effects based on party strategies}

\begin{figure*}[htbp]
\centering
\includegraphics[width=\textwidth]{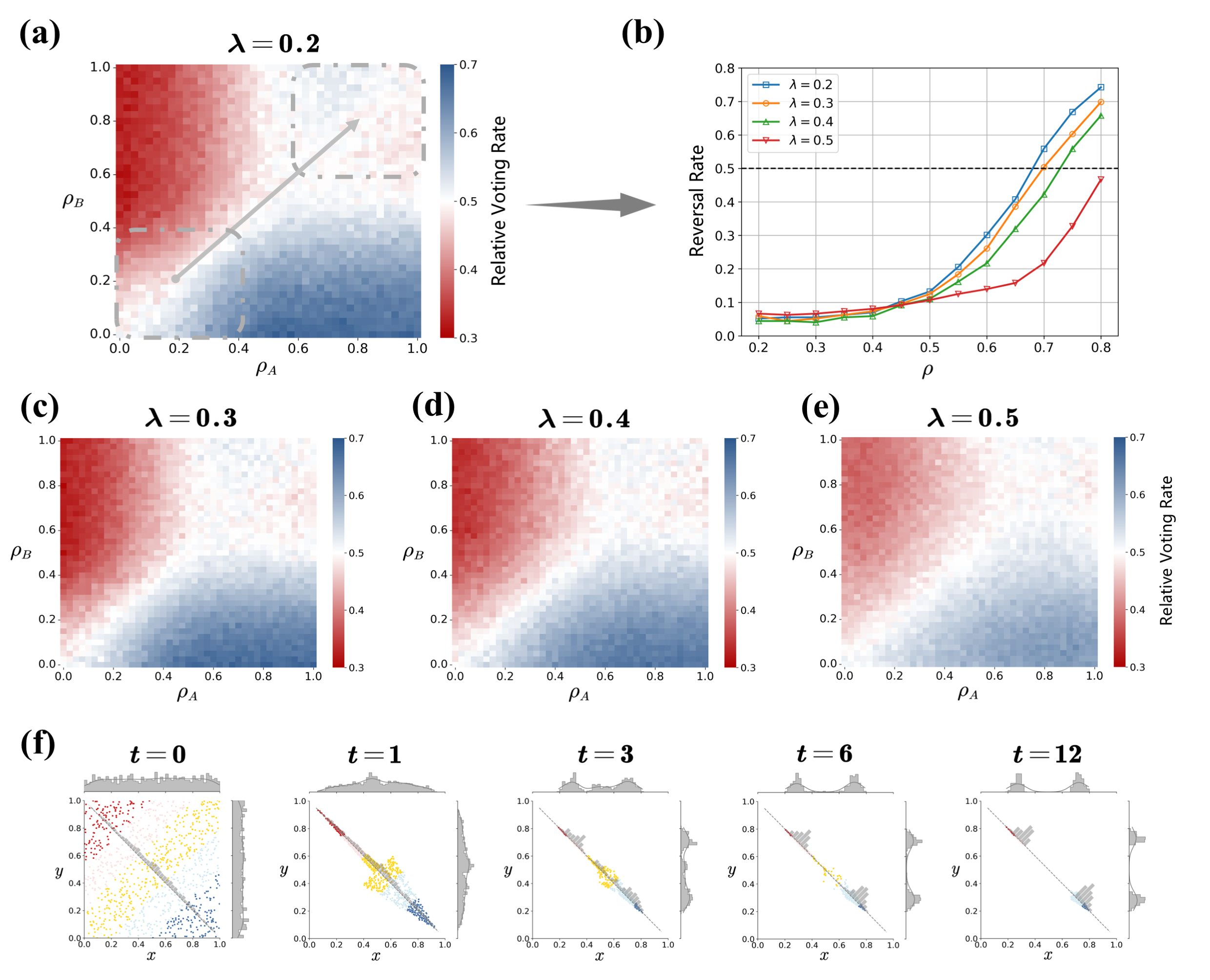}% Here is how to import EPS art
\caption{\label{fig:5}
Manipulation on swing voters and reversal of voting advantage. We apply heterogeneous antagonism effects to voters with different party preferences and show phase diagrams (a, c, d, e) at different stubbornness levels. Under low and moderate antagonism level, the party that imposes stronger antagonism can obtain higher vote support, and this advantage weakens as the group stubbornness level increases. When both parties create extremely strong antagonism, the party with weaker antagonism strategy may obtain more votes from swing voters and surprisingly achieve a reversal of voting advantage. We verify the existence of such reversal through Figure (b). We implement a square box with a length and width of 0.4 on Figure (a), letting its center slide along the diagonal direction. The horizontal axis is the antagonism level corresponding to the center of the rectangular box, and the vertical axis is the proportion of times the party with weaker antagonism reverse the voting advantage. Figure (f) shows the evolution snapshots of voting reversal under fixed parameters. Simulation results are averaged over 50 independent runs. Parameters:  simulations begin from an ER graph with $N=10^4$ and $\langle k \rangle=40$, $z_o=0.5$, $z_v=0.2$, and $\epsilon=0.2$. Other parameters: (a) $\lambda=0.2$. (c) $\lambda=0.3$. (d) $\lambda=0.4$. (e) $\lambda=0.5$. (f) $\lambda=0.2$, $\rho_A=0.6$, $\rho_B=0.9$.
}
\end{figure*}

Although the preceding analysis is restricted to symmetric and homogeneous parameter settings, our results have clearly suggested the possibility of altering election outcomes by manipulating swing voters through party-designed antagonistic interactions. In this section, we further examine how heterogeneous antagonism, exerted by different parties according to their own strategies, influences the election outcomes.

% Due to the model's inherent symmetry and homogeneous parameterization, group-level partisan opinions exhibit perfectly symmetric polarization, leading to nearly equal voting outcomes for the two parties---an outcome at odds with the asymmetric results of real-world presidential elections. Homogeneous antagonism, by design, fails to predict or regulate the differentiated outcomes of partisan competition. To address this, we investigate heterogeneous antagonism tailored to party preferences, enabling nuanced modeling of asymmetric manipulation of swingers.

% The sign of voters' opinion difference between two parties can serve as an indicator of their partisan inclinations and preferences. Leveraging this, we partition the voter population into two groups: those inclined towards Party A and those inclined towards Party B. Subsequently, we apply heterogeneous antagonism effects to these two groups, which are defined as follows:
We divide the population into two groups according to the voting tendency towards the two parties, i.e., the sign of opinion difference $z_i$, and apply heterogeneous antagonism effects to these two groups, which are defined as follows:
\begin{equation}
\rho_i = 
\begin{cases}
\rho_A, & \text{if } \text{sign}(z_i) = 1 \\
\rho_B, & \text{if } \text{sign}(z_i) = -1
\end{cases}
\end{equation}

In addition, we define the relative voting rate $p$ as the proportion of Party A voters among all actual voters that exclude the subgroup of neutrals.
Clearly, $p>0.5$ indicates that Party A holds a voting advantage, while $p<0.5$ implies that Party B secures the upper hand.

Figure~\ref{fig:5}(a) demonstrate that, under low and moderate level of antagonism, political parties with higher intensity of antagonism achieve a higher voting turnout. This can be attributed to the fact that increasing the strength of antagonism effect can effectively attract swing voters who lean towards a party and convert them into actual voters of that party. 
Nonetheless, when both parties adopt extreme campaigning strategies and exert strong antagonism, an intriguing phenomenon emerges: the party applying weaker antagonism attains a higher voting share, resulting in a dramatic reversal of voting advantage.

The robustness of these observations with respect to different levels of individual stubbornness $\lambda$ is verified in Figure~\ref{fig:5}(c-e).  We find that higher levels of stubbornness help the opinion system resist the influence of external antagonism, reducing the voting advantage of the dominant party. Of particular interest, as $\lambda$ increases, the area of the parameter space where reversal occurs decreases. 
To further quantitatively assess the probability and parameter region of reversal phenomenon, we implement a dynamics square window of size $0.4\times0.4$, whose center slides along the diagonal of the $\rho_a-\rho_b$ phase plane, and calculate the frequency of reversal within each window. Results in Figure 5(b) show that the reversal rate grows monotonically as the intensity of antagonism increases. Here we mark, with the threshold $p=0.5$, the regions in which reversal emerges relatively stably and with higher probability. We show that lower levels of stubbornness not only facilitate the onset of reversal, but also yield a higher reversal rate under identical antagonism intensities, thereby increasing the instability of electoral outcomes.

Further, to gain an in-depth understanding of the mechanism underlying the emergence of reversal, we characterize the temporal evolution process of group opinions under a set of fixed antagonism parameters, as illustrated in Figure~\ref{fig:5}(f).
We find that the reversal of voting advantage is rooted in the delicate transformation dynamics of the swing voters.
When Party B exerts more extreme antagonism, its leaners quickly side with Party B supporters, breaking the communication bridge between Party B and neutral voters. 
In contrast, Party A leaners gradually absorb these neutrals through sustained ideological interactions, which eventually induces the macro-scale reversal: the party adopting a weaker antagonistic strategy attracts more swing voters and, paradoxically, prevails in the fierce competition.

\subsection{Model performance on core networks during the 2020 U.S. presidential election}

\begin{figure*}[htbp]
\centering
\includegraphics[width=\textwidth]{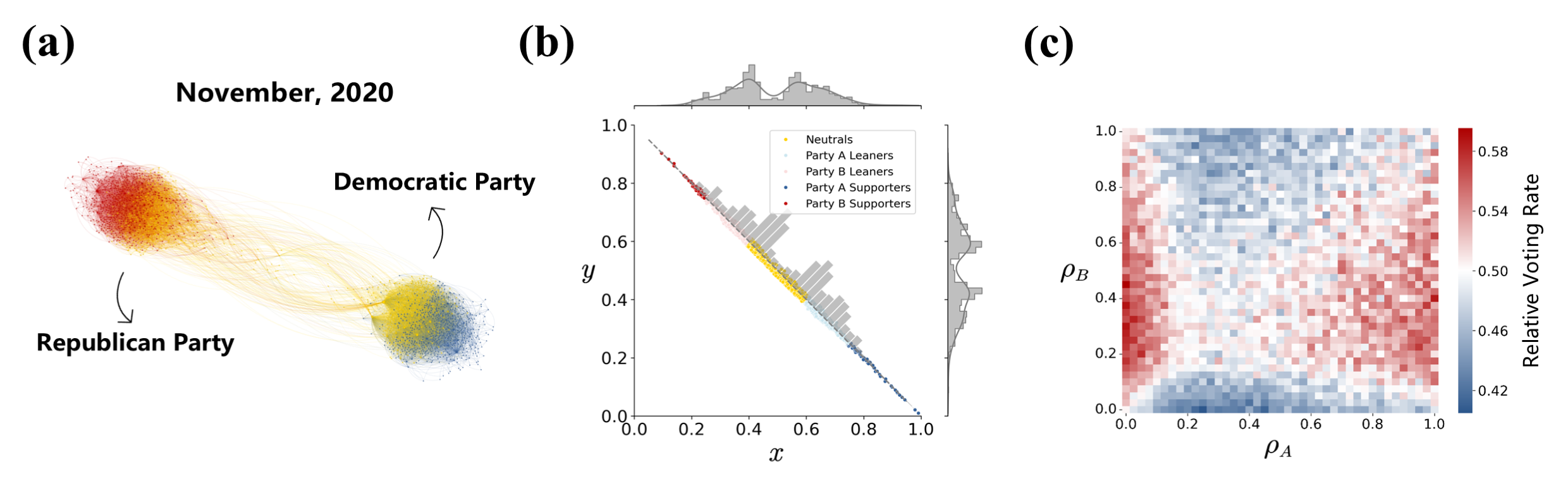}% Here is how to import EPS art
\caption{\label{fig:6}
Model performance in real-world networks and the robustness of key findings. We use Twitter (now X) data during the 2020 US election. In November 2020, we filter the retweet network using the $k$-core algorithm with $k=9$.
(a) The core retweet network after model simulations exhibits a symmetric echo-chamber structure, in which nodes represent user and edges denote effective retweet relationships. The blue, red, and yellow nodes represent Democratic voters, Republican voters, and neutral non-voters, respectively.
The final network contains 1981 nodes and 15737 edges.
(b) The opinion distributions of both parties and voting outcomes reveal a balanced polarization pattern between two parties. 
% (a) In October 2020, the core retweet network is an asymmetric echo-chamber structure after $k$-core filtering with $k=5$. The final network contains 2157 nodes and 11234 edges. (b)  The final network contains 1981 nodes and 15737 edges.
(c) We also verify the emergence of advantage reversal in the this network structure. Simulation results are averaged over 50 independent runs. Parameters: $z_o=0.5$, $z_v=0.2$, $\epsilon=0.2$, and $\lambda=0.2$. We fix: (b) $\rho=0.8$.
}
\end{figure*}

To validate the model performance under real-world topology, we apply our modeling framework to a representative political election. We utilize Twitter (now called X) data from the 2020 US presidential election spanning from November 1st to November 30th. 
% Retweets related to the election were gathered by setting specific filters containing selected keywords.
User IDs are modeled as network nodes, with directed retweet relationships serving as edges from source to target users, thereby constructing a data-driven network topology.

% partition the data by month and
Here, we filter it using the key word ``vaccine" to perform the typical case study.
Complying with previous studies, we extract the giant connected component and employ the $k$-core method to construct the core retweet network, deriving a clearer structure of the core groups while eliminating the interference from scattered and peripheral nodes (see data processing details in Appendix~\ref{app: 5}). 

Given that real-world retweet networks generally exhibit pronounced polarization, we identify two major partisan communities corresponding to the Democratic and Republican camps via the community detection algorithm. Users are thus categorized into two groups, and their initial opinions are assigned based on their community affiliations---distinct from the random initialization used in artificial networks before. Specifically, users in the Democratic community are set with $x(0)>y(0)$, whereas those in the Republican community satisfy $x(0)<y(0)$.

% In October, as the election approached its final stage, the retweet network displayed asymmetric polarization, with Democratic community occupying a substantially larger proportion. 

In November, as the election approached its final stage, the monthly retweet network displayed balanced polarization between the two parties based on the key word. 
We extract the core network using the $k$-core algorithm with $k=9$ and initialize opinions based on the detected communities. 
According to the model's final state---representing voting outcomes---users voting for the Democratic Party, Republican Party, and abstaining neutrals are visualized in blue, red, and yellow, respectively. 
The resulting network structure and opinion distributions, shown in Figure~\ref{fig:6}(a-b), indicate that the two partisan communities display symmetric structures and opinion distributions, with the neutral voter proportion accounting for 15-20\%.
These intermediary neutrals facilitate limited cross-chamber interaction, acting as bridges between the polarized voting coalitions.
% Democratic voters constitute more than half of the population, while neutrals account for approximately 20-25\%.
% Conversely, in November, after the election concluded, the retweet network exhibited balanced polarization between the two parties. Applying the $k$-core algorithm with $k=9$, we initialize and run the model on the extracted core network. As illustrated in Figure~\ref{fig:6}(b), the two partisan communities display symmetric structures and opinion distributions, with the neutral voter proportion accounting for 15-20\%.

Of particular interest, we validate the existence of reversal phenomenon in this case. 
In specific, we show the relative voting advantage in $\rho_A-\rho_B$ phase plane under the echo-chamber structure in Figure~\ref{fig:6}(c).
Compared with ER random networks, real retweet networks exhibit a even broader parameter region for generating advantage reversal. Notably, the pre-polarized network structure shows higher sensitivity to the modulation of antagonism effects, making it more prone to inducing large-scale reversals of voting advantage.

Overall, we demonstrate that in real election networks there is an even greater likelihood of voting advantage reversal induced by extreme antagonism, highlighting the delicate balance required to manipulate swing groups in actual elections.
The robustness of voting reversal regarding to various network topologies is shown in Appendix~\ref{app: 4}

\section{CONCLUSIONS AND DISCUSSIONS}
Under the furious circumstances of contemporary presidential elections, the behavior of swing voters---those navigating the ideological spectrum between polarized parties---represents a pivotal nexus where democratic outcomes are determined. These individuals embody the intricate interplay of ideological flexibility, social influence, and external political campaigns. Yet existing frameworks often oversimplifies them as ``undecided'' and  anchors in one-dimensional assumptions of binary opposition (support for one party as inherent opposition to another), failing to capture the multidimensionality of their attitudes. Swing voters may hold issue-specific preferences that defy strict partisan alignment or adopt neutrality as a strategic response to hostile political climates, a complexity that traditional models obscure.
Moreover, these frameworks overlook the non-trivial coupling effects of top-down campaigns and bottom-up opinion interactions.
These oversights have limited our ability to explain how these groups solidify into decisive blocs, retreat into abstention, or even trigger electoral shocks---dynamics that define modern democratic elections.

This study addresses this gap with a sociologically rich framework that reimagines political attitude as a two-dimensional space, where stances for opposing parties evolve not in lockstep but through dynamic interactions between campaign strategies and psychological dispositions. By examining partisan antagonism---the degree to which campaigns frame politics as an adversarial ``us vs. them'' contest, the research reveals several interconnected insights with profound societal relevance. 
Firstly, the external antagonism effect systematically erodes the diversity of ideological interactions while enhances voting mobilization, leading to the formation of two-dimensional echo chambers. This process not only mobilizes neutrals into partisan blocs but also creates self-reinforcing feedback loops where voters are increasingly insulated from opposing views, mirroring empirical patterns in social media networks \cite{cinelli2021echo,cota2019quantifying,flamino2023political}. 
Secondly, the study uncovers a complex interaction between top-down antagonistic strategies and bottom-up individual stubbornness, which exerts a nonlinear impact on the emergence and evolution of the swing voters. 
Last but not least, we find a nuanced relationship between competitive antagonism and electoral outcomes: Under weak to moderate antagonism, the party with more intense antagonistic strategies secures a voting advantage by mobilizing more swing voters. Paradoxically, when both parties deploy extreme antagonism, the resultant ideological competition can backfire, triggering advantage reversal where the party with weaker antagonism gains more support from neutrals. 

% 等数据实验结束后大改
Analysis of retweet networks during 2020 US presidential election reveals that real-world echo chambers---characterized by homogeneous partisan interaction and limited cross-ideological engagement---create fertile ground for the model's predicted advantage reversal under extreme antagonism. In such polarized environments, swing voters trapped in information silos may defect to the less aggressive party as a psychological response to cognitive overload \cite{haynes2010cognitive,metag2023too}. This phenomenon, rooted in sociological patterns of homophily and psychological mechanisms of cognitive dissonance reduction, highlights how digital segregation amplifies the risk of unpredictable electoral shifts \cite{tornberg2022digital,antelmi2025characterizing}. 

Our framework has certain limitations. 
We adopt the static network structure, ignoring the evolution of the network itself caused by widely used AI recommendation algorithms that can be considered in future research. Additionally, to simplify our model and emphasize the coupling influence from top to bottom and bottom to top, we only consider the heterogeneity of external manipulation and the dynamic interactions among different types of voters, without taking into account the heterogeneity of individual stubbornness and open-mindedness. Future research may attribute the heterogeneity of individual characteristics based on real data.

In an age of increasing polarization and democratic strain driven by the growing complexity of social media, these findings highlight the need to reconsider the roles of political strategies and their impact on civic life \cite{aral2019protecting}. We show the possibility of balancing clarity with ambiguity, using antagonism to define choices without destroying the informational diversity that enables meaningful engagement. This work reminds us that the future of democracy hinges not on the power of partisan extremes, but on the capacity to honor the complexity of swing voter dynamics---a sociological reality that holds the key to inclusive governance and resilient democratic cultures. 
Furthermore, although based on election contexts, our modeling framework is, in fact, applicable to studying the evolution and manipulation of opinions in all polarized environments, such as vaccines, gun control, abortion, immigration, and other issues.

\begin{acknowledgments}
This work is supported by National Science and Technology Major Project (2022ZD0116800), Program of National Natural Science Foundation of China (62141605, 12425114, 12201026, 12301305), the Fundamental Research Funds for the Central Universities, Beijing Natural Science Foundation (Z230001), and Beijing Advanced Innovation Center for Future Blockchain and Privacy Computing.
\end{acknowledgments}

\appendix

\section{CONVEXITY AND GLOBAL MINIMUM OF THE OPINION UPDATING FUNCTION}
\label{app: 1}
In this appendix, we provide a comprehensive analysis of the convexity of the simplified opinion updating function $\mathcal{L}(x_i,y_i)$ and derive its unique global minimum. Demonstrating the convexity and finding the global minimum is crucial as it justifies the feasibility and correctness of using convex optimization methods for opinion updating.

\subsection{Convexity proof of the function}

The function $\mathcal{L}(x_i,y_i)$ is defined as:
\begin{equation}
\begin{aligned}
\mathcal{L}(x_i,y_i) =  (1 - \rho) \big[ \lambda \cdot (x_i - x_i(0))^2 + (1 - \lambda)(x_i - \bar{x}_i(t))^2 + \\
\lambda \cdot (y_i - y_i(0))^2 + (1 - \lambda)(y_i - \bar{y}_i(t))^2 \big] +\rho (x_i + y_i - 1)^2
\end{aligned}
\end{equation}
where $\lambda\in[0,1]$ represents a weighting parameter related to the influence of initial opinions and current neighborhood-averaged opinions, and $\rho\in[0,1]$ is a parameter governing the strength of the antagonism.

We first calculate the first-order partial derivatives. The partial derivative with respect to $x_i$ is:

\begin{equation}
\begin{aligned}
\frac{\partial\mathcal{L}}{\partial x_i}&=2[x_i-(1 - \rho)\lambda x_i(0)-(1 - \rho)(1 - \lambda)\bar{x}_i(t)+\rho y_i-\rho]
\end{aligned}
\end{equation}

The partial derivative with respect to $y_i$ is:
\begin{equation}
\begin{aligned}
\frac{\partial\mathcal{L}}{\partial y_i}&=2[y_i-(1 - \rho)\lambda y_i(0)-(1 - \rho)(1 - \lambda)\bar{y}_i(t)+\rho x_i-\rho]\\
\end{aligned}
\end{equation}

Next, we find the second-order partial derivatives:
\begin{subequations}
\begin{eqnarray}
\frac{\partial^2\mathcal{L}}{\partial x_i^2}=2\left[(1 - \rho)+\rho\right]=2, \label{appa}
\\
\frac{\partial^2\mathcal{L}}{\partial y_i^2}=2\left[(1 - \rho)+\rho\right]=2, \label{appb}
\\
\frac{\partial^2\mathcal{L}}{\partial x_i\partial y_i}=\frac{\partial^2\mathcal{L}}{\partial y_i\partial x_i}=2\rho. \label{appc}
\end{eqnarray}
\end{subequations}

The Hessian matrix $H$ of the function $\mathcal{L}(x_i,y_i)$ is:
\begin{equation}
H=\begin{bmatrix}
\frac{\partial^2\mathcal{L}}{\partial x_i^2}&\frac{\partial^2\mathcal{L}}{\partial x_i\partial y_i} \\
\frac{\partial^2\mathcal{L}}{\partial y_i\partial x_i}&\frac{\partial^2\mathcal{L}}{\partial y_i^2}
\end{bmatrix}=\begin{bmatrix}
2&2\rho\\
2\rho&2
\end{bmatrix}
\end{equation}

For a $2\times2$ matrix $A=\begin{bmatrix}a&b\\c&d\end{bmatrix}$ to be semi-positive definite, we need $a\geq0,d\geq0$, and $ad-bc\geq0$. In matrix $H$, $a=2\geq0$, $d=2\geq0$, and $ad-bc=4-4\rho^2=4(1-\rho^2)$. Since $\rho\in[0,1]$, then $1-\rho^2\geq0$, so $ad-bc\geq0$. Thus, the Hessian matrix $H$ is semi-positive definite, and the function $\mathcal{L}(x_i,y_i)$ is convex. This convexity justifies the use of convex optimization in our opinion updating model.

\subsection{Finding the unique global minimum}

Now we can proceed to find its unique global minimum, which occurs at the point where the first-order partial derivatives are equal to zeros. That is, we need to solve the following system of equations:
\begin{equation}
\frac{\partial\mathcal{L}}{\partial x_i} = \frac{\partial\mathcal{L}}{\partial y_i}= 0
\end{equation}

In the case when $\rho\ne1$, we solve the system of equations for $x_i,y_i$:

\begin{equation}
\begin{aligned}
\hat{x_i} &= \frac{(1-\rho)[\lambda x_i(0)+(1 - \lambda) \bar{x}_i] + \rho}{1 - \rho^2} \\&- \frac{ \rho(1-\rho) [\lambda y_i(0) + (1 - \lambda) \bar{y}_i] + \rho^2}{1 - \rho^2}
\end{aligned}
\end{equation}

\begin{equation}
\begin{aligned}
\hat{y_i} &= \frac{(1-\rho)[\lambda y_i(0)+(1 - \lambda) \bar{y}_i] + \rho}{1 - \rho^2} \\&-\frac{ \rho(1-\rho) [\lambda x_i(0) + (1 - \lambda) \bar{x}_i] + \rho^2}{1 - \rho^2}
\end{aligned}
\end{equation}

which can be simplified to:
\begin{equation}
\hat{x_i} = \frac{[\lambda x_i(0)+(1 - \lambda) \bar{x}_i] - \rho [\lambda y_i(0) + (1 - \lambda) \bar{y}_i] + \rho}{1 + \rho}
\end{equation}

\begin{equation}
\hat{y_i} = \frac{[\lambda y_i(0) + (1 - \lambda) \bar{y}_i] - \rho[\lambda x_i(0) + (1 - \lambda) \bar{x}_i] + \rho}{1 + \rho}
\end{equation}

When $\rho=1$, the optimization objective  $\mathcal{L}(x_i,y_i)$ simplifies to a constant value. Specifically, as shown in the previous derivations, when substituting $\rho=1$ into $\mathcal{L}(x_i,y_i)$, we get $\mathcal{L}(x_i,y_i) = (x_i+y_i-1)^2$. And from $\frac{\partial\mathcal{L}}{\partial x_i}=\frac{\partial\mathcal{L}}{\partial y_i}=2x_i+2y_i-2=0$, we have $x_i+y_i=1$ making $\mathcal{L}(x_i,y_i)=0$.

In such a scenario, for the sake of consistency in the overall solution framework and without causing any contradictions in the theoretical context, we can artificially define the solution of the optimization problem when $\rho=1$ to be identical to the form obtained in the general case when $\rho\ne1$.

Consequently, in summary, by solving for the global minimum of the convex objective function, we obtain the following dynamics equations:

\begin{equation}
\begin{aligned}
x_i(t+1) &= \frac{[\lambda x_i(0) + (1 - \lambda) \bar{x}_i] - \rho [\lambda y_i(0) + (1 - \lambda) \bar{y}_i] + \rho}{1 + \rho} \\
y_i(t+1) &= \frac{[\lambda y_i(0) + (1 - \lambda) \bar{y}_i] - \rho [\lambda x_i(0) + (1 - \lambda) \bar{x}_i] + \rho}{1 + \rho}
\end{aligned}
\end{equation}
where $\rho\in[0,1]$.

\section{HETEROGENEOUS ANTAGONISM AT LOW OPEN-MINDEDNESS}
\label{app: 2}

\begin{figure*}[htbp]
\centering
\includegraphics[width=\textwidth]{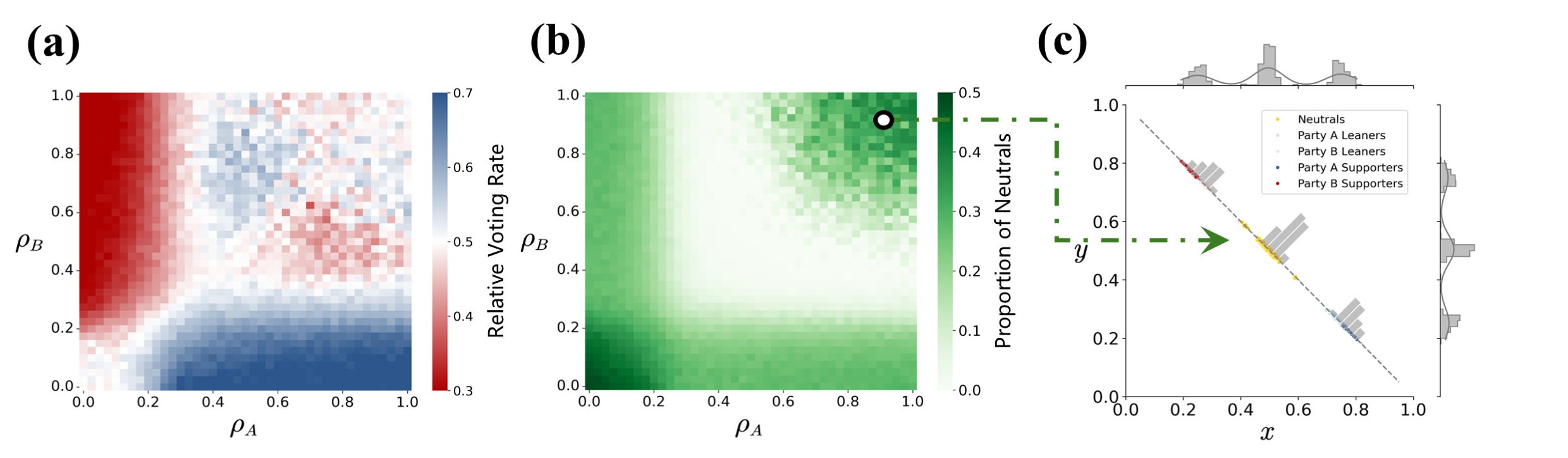}% Here is how to import EPS art
\caption{\label{fig:7} 
The relative voting rate (a) and the number of neutrals (b) under heterogeneous antagonism at low open-mindedness. When both parties impose extremely intense levels of antagonism (c), a neutral group may emerge due to the tripolarization of ideological stances. Under such circumstances, neither the opposing party nor the weaker party manages to garner additional voting support from this middle group and the phenomenon of advantage reversal fails to materialize. 
Simulation results are averaged over 50 independent runs. Parameters: Simulations in (a) and (b) begin from an ER graph with $N=10^4$ and $\langle k \rangle=40$, $z_o=0.5$, $z_v=0.2$, $\epsilon=0.15$, and $\lambda=0.2$. We change (c) $\rho_A=\rho_B=0.9$.
}
\end{figure*}

In the main text, we demonstrate that heterogeneous antagonism under moderate interaction open-mindedness drives the establishment and potential reversal of partisan voting advantages. Here, we characterize the dynamics under a low open-mindedness, where interaction constraints intensify  ideological segregation, leading to distinct behavioral regimes compared to the moderate-threshold case.

A low $\epsilon$ imposes strict limits on cross-ideological engagement, confining voters to highly homophilous interactions---primarily with neighbors sharing nearly identical opinions. This structural isolation solidifies preexisting opinions, inhibiting the diffusion of opposing opinions and fostering the emergence of tripolarized partisan configurations: cohesive blocs of Party A voters, Party B voters, and a sizeable neutral bloc that remains ideologically detached from both extremes (Fig.~\ref{fig:7}(c)).  

Under moderate open-mindedness, elevating the level of heterogeneous antagonism effectively drives the conversion of swing voters to a specific party, thereby boosting its voter turnout. Compared with Figure~\ref{fig:5}(a), in regimes with severely constrained interactions as illustrated in Figure~\ref{fig:7}(a,b), heightened antagonism exacerbates ideological polarization, amplifying the dominant party's voting advantage through the accelerated assimilation of marginally aligned individuals while curbing cross-ideological exchanges that might otherwise moderate partisan extremes.

Notably, when both parties deploy strong antagonism ($\rho_A,\rho_B>0.5$) under low $\epsilon$, a complex threshold phase transition emerges. The elimination of neutrals and the reversal of voting advantage, observed in moderate-threshold scenarios, becomes diverse at low $\epsilon$:

\begin{itemize}
    \item Reversal under near-polarization: When the neutral group is nearly eliminated ($\rho_A,\rho_B$ both relatively high), the weak party can still attract residual swing voters, leading to reversals similar to the moderate-threshold case.

    \item Stable dominance with persistent neutrality ($\rho_A,\rho_B$ both extremely high): Conversely, if a large neutral group persists due to extreme interaction restrictions, the dominant party's antagonism effectively freeze the electoral landscape, as neutrals neither participate nor shift allegiances, cementing the status quo.
\end{itemize}

Mechanistically, the ability to reverse voting advantages hinges on the neutral bloc's connectivity to partisan groups: low $\epsilon$ reduces the neutral bloc's susceptibility to extreme influence, making it a passive bystander rather than a swing constituency. Thus, advantage reversal occurs only when interaction constraints are sufficiently relaxed to allow limited cross-ideological signaling, even at low levels, enabling the weak party to gradually attract neutral voters through indirect ideological cues.

In summary, low open-mindedeness exacerbate the structural barriers to opinion diffusion, transforming heterogeneous antagonism from a driver of swing voter mobilization into a stabilizer of polarized or tripolarized states.  These findings underscore the critical role of interaction topology in mediating the efficacy of antagonistic strategies, revealing how even subtle changes in engagement rules can fundamentally alter the dynamics of partisan competition and electoral outcomes.

\section{MODEL ROBUSTNESS TO STRUCTURAL PARAMETERS}
\label{app: 3}
In our framework, the mesoscale multi-cognitive classification of voters and intergroup interaction dynamics are governed by two critical structure parameters: opinion difference thresholds ($z_o,z_v$), and influence weights ($\omega$). Here, we assess the model's robustness to variations in these parameters, which define the boundaries of partisan categorization and the strength of intergroup influence, respectively.

\subsection{Robustness to multi-cognitive thresholds}
\label{app: 3a}

\begin{figure*}[htbp]
\centering
\includegraphics[width=\textwidth]{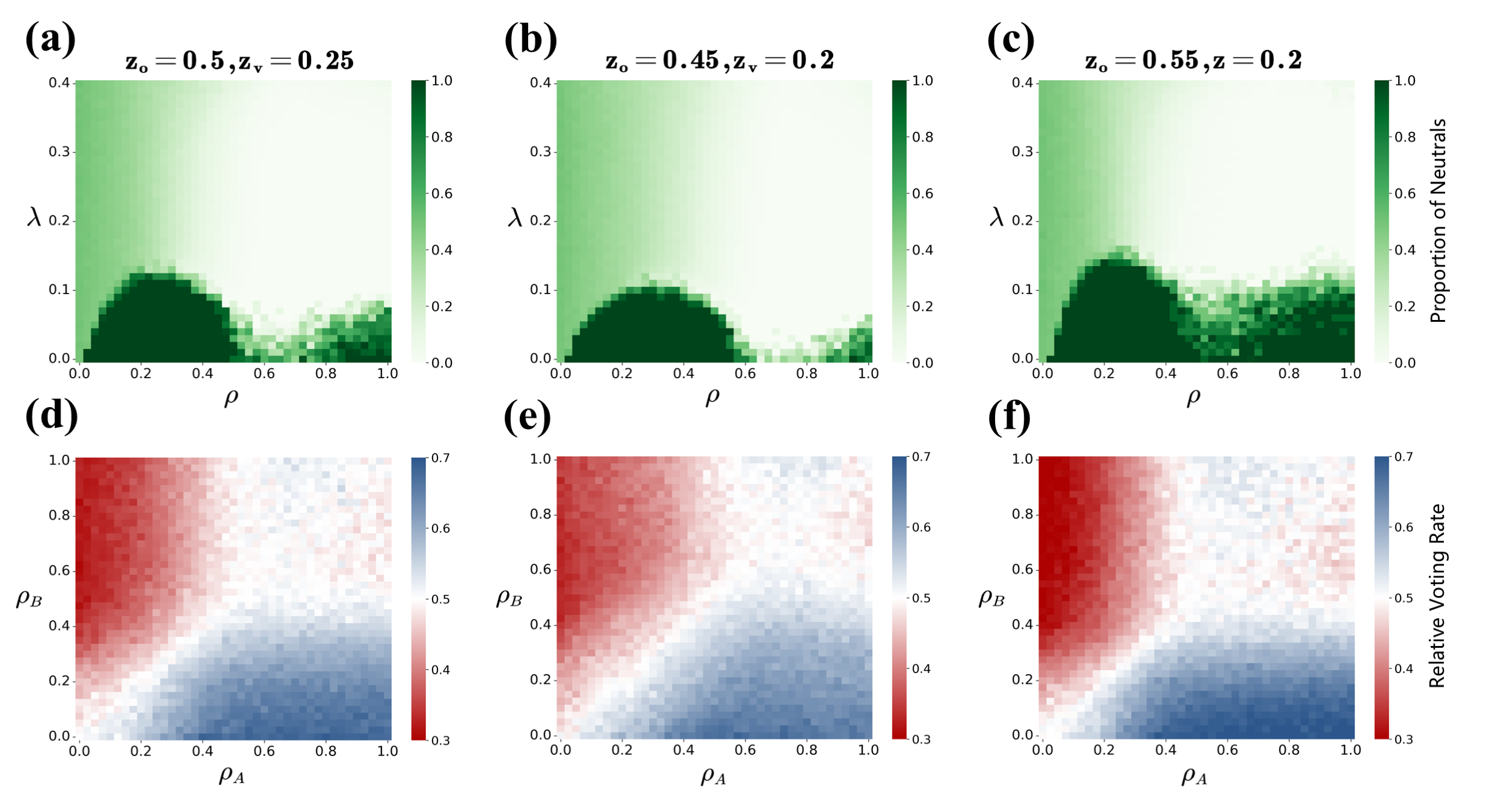}% Here is how to import EPS art
\caption{\label{fig:8} 
Effects of nonlinear coupling between stubbornness and antagonism levels (a-c) and heterogeneous antagonism levels (d-e) under different multi-cognitive thresholds. Results largely resemble findings from the original setting. The fine-tuning of $z_o$ results in the shift of partisan polarization boundary. Simulation results are averaged over 50 independent runs. Parameters: simulations begin from an ER graph with $N=10^4$, $\langle k \rangle=40$ and $\epsilon=0.2$. We change (a,d) $z_o=0.5,z_v=0.25$. (b,e) $z_o=0.45,z_v=0.2$. (c,f) $z_o=0.55,z_v=0.2$. (d,e,f) $\lambda=0.2$.
}
\end{figure*}

\begin{figure*}[htbp]
\centering
\includegraphics[width=\textwidth]{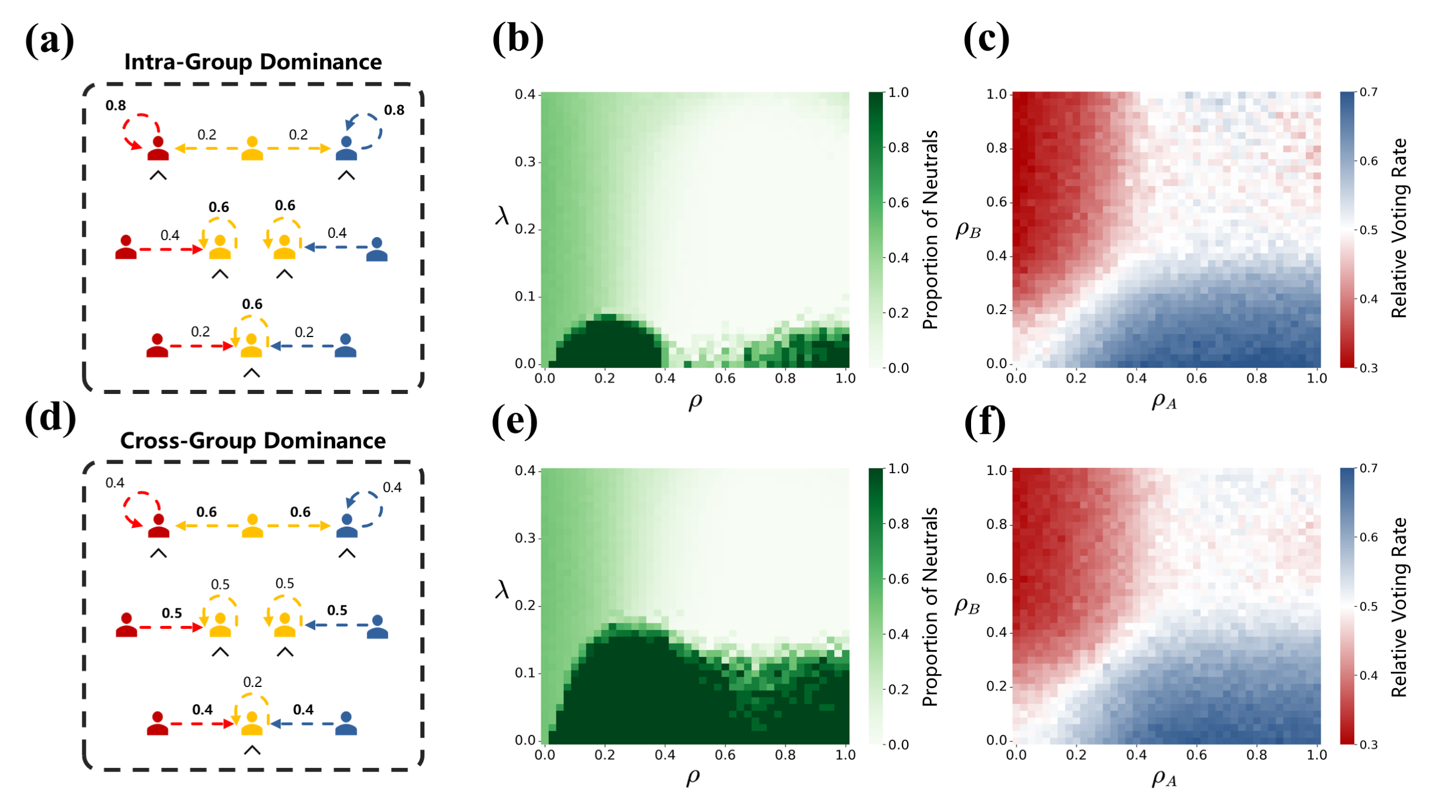}% Here is how to import EPS art
\caption{\label{fig:9}
Effects of nonlinear coupling between stubbornness and antagonism levels and heterogeneous antagonism levels under two sets of influence weights. 
We run simulations under the influence weights with intra-group dominance (a-c) and cross-group dominance (d-f).
Results largely resemble findings from the original setting. Simulation results are averaged over 50 independent runs. Parameters: simulations begin from an ER graph with $N=10^4$ and $\langle k \rangle=40$, $z_o=0.5$, $z_v=0.2$, and $\epsilon=0.2$. We fix (c,f) $\lambda=0.2$. 
}
\end{figure*}

The thresholds $z_o$ and $z_v$ partition the voter population into five behavioral classes by quantifying the salience of attitudinal polarization. To validate robustness, we systematically vary $z_o\in\{0.45,0.55\}$ and $z_v\in\{0.15, 0.25\}$, spanning realistic ranges of attitudinal ambiguity.

Results in Figure~\ref{fig:8} show that while the size of mesoscale groups changes with threshold settings, the qualitative dynamics of opinion polarization and voting outcome tripolarization remain invariant. 

Specifically, the inner threshold $z_v$ defines the voting boundary but does not alter the mesoscale interaction mechanisms, which are governed by the outer threshold $z_o$. While $z_v$ influences the size of the neutral population, the nonlinear coupling between stubbornness and antagonism and the establishment and reversal of heterogeneous antagonism advantage persist across $z_v$ fine-tunings, confirming in Figure~\ref{fig:8}(a,d) that the model's core behavioral predictions are insensitive to the precise definition of voting boundaries. 

The outer threshold $z_o$, by contrast, directly modulates the mesoscale composition of voter groups:
\begin{itemize}
    \item Partisan Polarization Boundary.--- Contraction of $z_o$ ($z_o \downarrow$) narrows the boundary for strong partisans, expanding the sizes of Party A/B supporters at the expense of swing voters. This shift reduces the pool of malleable leaners, inhibiting cross-ideological interactions and reinforcing intra-party homogeneity. Consequentially, the system exhibits enhanced polarization stability across a wider parameter space (Fig.~\ref{fig:8}(b)), as the diminished swing population weakens the moderating effect of intergroup influence. Conversely, expanding $z_o$ ($z_o\uparrow$) widens the threshold for strong partisans, increasing the proportion of swing voters and fostering broader ideological engagement (Fig.~\ref{fig:8}(c)). Larger leaner groups facilitate cross-cutting interactions and consensus-building mechanisms, making the model more prone to moderate, non-polarized outcomes under higher $z_o$.

    \item Impactions for Partisan Advantage Reversal.--- Lower $z_o$ values, by shrinking swing voter pools and intensifying inter-party segregation, impose structural barriers to advantage reversal under extreme antagonism (Fig.~\ref{fig:8}(e)). The reduced availability of ideologically flexible neutrals and leaners limits the weak party's ability to attract support, even when confronting strong antagonism from the dominant party. Conversely, by preserving a sizeable swing population, higher $z_o$ maintains the conditions for threshold-dependent reversals, as moderate leaners remain susceptible to strategic antagonism and intergroup influence (Fig.~\ref{fig:8}(f)).
\end{itemize}

In summary, while $z_v$ governs the magnitude of neutrals without altering core dynamics, $z_o$ acts as a critical lever for balancing polarization and flexibility in partisan competition. The model's qualitative predictions---including tripolarization, antagonism-driven mobilization, and advantage reversal---remain robust to $z_o$ adjustments, as the underlying mechanisms of intergroup interaction and convex opinion updating are preserved across reasonable threshold configurations. This robustness ensures the model's applicability to diverse political contexts, from highly polarized to moderately pluralistic electoral systems.

\subsection{Robustness to influence weights}
\label{app: 3b}

\begin{figure*}[htbp]
\centering
\includegraphics[width=\textwidth]{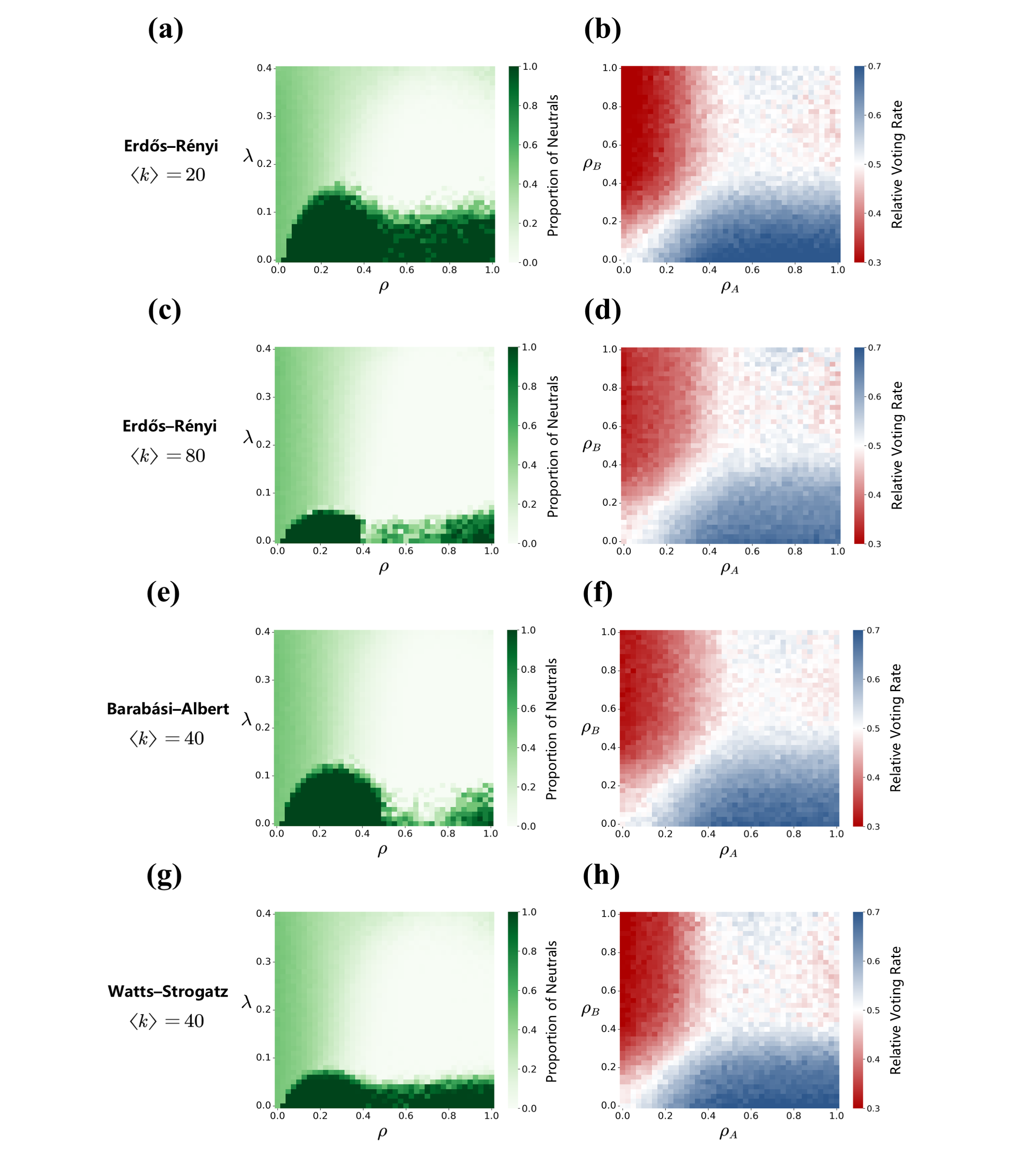}% Here is how to import EPS art
\caption{\label{fig:10} 
Effects of nonlinear coupling between stubbornness and antagonism levels and heterogeneous antagonism levels under four different network topologies. 
We run simulations under ER random networks with $\langle k\rangle=20$ (a-b) and $\langle k\rangle=80$ (c-d), under a BA network with $\langle k\rangle=40$ (e-f) and under a WS network with $\langle k\rangle=40$ (g-h), respectively.
Results largely resemble findings from the original setting. Simulation results are averaged over 50 independent runs. Parameters: simulations begin from networks with $N=10^4$, $z_o=0.5$, $z_v=0.2$, and $\epsilon=0.2$. We fix (b,d,f,h) $\lambda=0.2$. 
}
\end{figure*}

The influence weights $w$, which parameterized the strength of intergroup interactions (e.g. $w_{AX},w_{CX}$ in Table ~\ref{tab:table1}), govern how mesoscale group opinions propagate through the network. We test robustness by perturbing these weights within theoretically justified ranges. 

In light of the equilibrium of group interaction, we have meticulously configured two distinct sets of influence weights shown in Figure~\ref{fig:9}(a,d): one favoring intra-group interaction dominance and the other emphasizing cross-group interaction dominance. Under the regime of intra-group interaction dominance, voters belonging to the same category assign a higher influence weight when accepting each other's opinions, while being less significantly influenced by opinions of voters from other categories. Conversely, the setting of cross-group interaction dominance presents an essentially opposite scenario.

Despite the alteration in the opinion convergence rate, results indicate that the model consistently exhibits stable macroscale outcomes. The observed differences in nonlinear phenomena are merely of an appropriate quantitative nature, without affecting the the overall stability and integrity of the model's macroscopic manifestations.

\begin{itemize}
   \item Partisan Polarization Boundary.---When the interaction of voters' opinions is predominantly shaped by intra-group forces, the exchange of opinions between the extreme partisan supporters and the swing groups is severely restricted. This constraint poses a significant obstacle to the emergence of group consensus at the macro level (Fig.~\ref{fig:9}(b)). In the phase diagram depicting the relationship between the stubbornness level and the antagonism level, polarization is prevalent and stable. Group consensus can only be achieved under conditions of extremely flexible initial preferences. In contrast, when greater consideration is given to the opinions of different types of voters, large-scale interactions become conducive to the generation of group consensus (Fig.~\ref{fig:9}(e)).

    \item Impactions for Partisan Advantage Reversal.---When the interaction of voters' opinions is mainly governed by intra-group influence, the limited exchange of opinions actually facilitates the establishment of partisan advantages at a low antagonism level. However, as the antagonism level rises, the dominant party experiences a more rapid disconnection from the neutral group, thereby triggering a reversal of the voting advantage (Fig.~\ref{fig:9}(c)). On the contrary, in an environment that promotes cross-group interaction, the establishment and reversal of partisan advantages occur to a lesser degree (Fig.~\ref{fig:9}(f)).
\end{itemize}

Overall, the model demonstrates unwavering robustness in the face of structural parameter variations. Such parameter changes do not impinge upon the model's qualitative outcomes in any substantial way. Instead, the impact is confined to minor quantitative discrepancies, leaving the model's fundamental qualitative characteristic intact.

\section{MODEL ROBUSTNESS TO NETWORK TOPOLOGY}
\label{app: 4}

In the main text, the model network is initialized as an Erdős–Rényi (ER) random graph with $N=10^4$ and connection probability $p=0.004$, corresponding to an average degree of $\langle k\rangle=40$. To examine the robustness of our results with respect to network topology, we further consider ER networks with $p=0.002$ ($\langle k\rangle=20$) and $p=0.008$ ($\langle k\rangle=80$), as well as a Barabási–Albert (BA) network with $m=40$ ($\langle k\rangle=40$) and a Watts–Strogatz (WS) network with $k=40$ and rewiring probability $p=0.1$ ($\langle k\rangle=40$).
The results in Figure~\ref{fig:10} indicate that, despite the variations in the network topology and the average degree, the qualitative dynamics and macroscale results remain invariant. 

The average degree directly determines the overall connectivity of the network, thereby influencing the threshold conditions for partisan polarization. Within the same ER topology, varying the average degree reveals that higher connectivity leads to a broader onset of polarization, as evidenced by the shift of the polarization boundary in Figure~\ref{fig:10}(a,c).
When polarization occurs under moderate and high stubbornness, the antagonism effect on promoting the voting participation of swing voters remains comparable. In cases where the two parties engage in competitive antagonism, no significant difference is observed in the manipulation of swing voters or in the resulting voting outcomes in Figure~\ref{fig:10}(b,d).

We further verify that these outcomes still hold across heterogeneous and clustered network structures by replacing the ER topology with BA and WS networks. 
In particular, under the BA network with the same average degree as the ER network, we observe the nonlinear coupling pattern in Figure~\ref{fig:10}(e) almost identical to that shown in Figure~\ref{fig:4}(b). This consistency strongly demonstrates the model robustness against variations in the network topology. In contrast, the $\lambda-\rho$ phase diagram of the WS network in Figure~\ref{fig:10}(g) is similar to that of the ER network with $\langle k\rangle=80$.
Furthermore, we find that the qualitative establishment and reversal of voting advantages under antagonistic strategies remain robust, with only slight differences in the extent of reversal, as shown in Figure~\ref{fig:10}(f,h).

These results demonstrate that the core dynamics of the model are structurally robust across both homogeneous and heterogeneous topologies, with only slight quantitative shifts under different degree distributions and clustering coefficients.

\section{TWITTER DATA PROCESSING}
\label{app: 5}

All original Twitter datasets are publicly available as Twitter IDs \cite{liu2025github}. We first download the tweets using Twitter's API and obtain user ID (nodes), retweeting relationship (edges), and political orientation of tweets by month. Due to the computational constraints, we do not simulate the full network. Instead, we filter the networks with several key words (e.g. vaccine, abortion and covid).
For visualization purposes, network layouts are generated using the Force Atlas 2 algorithm implemented in \textsc{gephi}, with node sizes weighted by their retweet frequencies.
To extract the core structure of the network for our analysis, we apply the $k$-core algorithm to remove nodes with fewer retweet relationships, which leads to a more simplified network.
Furthermore, by quantifying the political leanings reflected in partial retweets, we infer the political preferences of the core nodes, thereby characterizing partisan orientation to each community.
% The sample datasets we use in this paper and the data-processing codes (written in \textsc{PYTHON3}) can be obtained at ...

All the network visualizations in this paper are generated using the Force Atlas 2 layout algorithm implemented in \textsc{gephi}. The color of each node is determined by the voting outcomes obtained from model simulations.

% The \nocite command causes all entries in a bibliography to be printed out
% whether or not they are actually referenced in the text. This is appropriate
% for the sample file to show the different styles of references, but authors
% most likely will not want to use it.

% \nocite{*}
% \bibliography{apssamp}% Produces the bibliography via BibTeX.

%apsrev4-2.bst 2019-01-14 (MD) hand-edited version of apsrev4-1.bst
%Control: key (0)
%Control: author (8) initials jnrlst
%Control: editor formatted (1) identically to author
%Control: production of article title (0) allowed
%Control: page (0) single
%Control: year (1) truncated
%Control: production of eprint (0) enabled
%

\end{document}